# Efficiency Loss in a Network Resource Allocation Game: The Case of Elastic Supply

Ramesh Johari (`rjohari@stanford.edu`)
Shie Mannor (`shie@mit.edu`)
John N. Tsitsiklis (`jnt@mit.edu`)

July 1, 2018

**Abstract**

We consider a resource allocation problem where individual users wish to send data across a network to maximize their utility, and a cost is incurred at each link that depends on the total rate sent through the link. It is known that as long as users do not anticipate the effect of their actions on prices, a simple proportional pricing mechanism can maximize the sum of users' utilities minus the cost (called aggregate surplus). Continuing previous efforts to quantify the effects of selfish behavior in network pricing mechanisms, we consider the possibility that users anticipate the effect of their actions on link prices. Under the assumption that the links' marginal cost functions are convex, we establish existence of a Nash equilibrium. We show that the aggregate surplus at a Nash equilibrium is no worse than a factor of $4\sqrt{2} - 5$ times the optimal aggregate surplus; thus, the efficiency loss when users are selfish is no more than approximately 34%.

The current Internet is used by a widely heterogeneous population of users; not only are different types of traffic sharing the same network, but different end users place different values on their perceived network performance. This has led to a surge of interest in *congestion pricing*, where the network is treated as a market, and prices are set to mediate demand and supply of network resources; see, e.g., [5, 9].

We investigate a specific price mechanism considered by Kelly et al. in [16] (motivated by the proposal made in [14]). For simplicity let us first consider the special case of a single link; in this case the mechanism works as follows. Each user submits a bid, or total *willingness-to-pay*, to the link manager. This represents the total amount the user expects to pay. The link manager then chooses a total rate and price such that the product of price and rate is equal to the sum of the bids, and the price is equal to marginal cost; note, in particular, that the supply of the link is *elastic*, i.e., it is not fixed in advance. Finally, each user receives a fraction of the allocated rate in proportion to their bid. It is shown in [16] that if users do not anticipate the effect of their bid on the price, such a scheme maximizes the sum of users' utilities minus the cost of the total allocated rate, known as the *aggregate surplus* (see [20], Chapter 10).

The pricing mechanism of [16] takes as input the bids of the users, and produces as output the price of the link, and the resulting rate allocation to the users. Kelly et al. [16] continue on to discuss distributed algorithms for implementation of this market-clearing process: given the bids of the users, the authors present two algorithms which converge to the market-clearing price and rate allocation. Indeed, much of the interest in this market mechanism stems from its desirable properties as a decentralized system, including both stability and scalability. For details, we refer the reader to [11, 13, 27, 28].

One important interpretation of the price given to users in the algorithms of [16] is that it can provide early notification of congestion. Building on the Explicit Congestion Notification (ECN) proposal [22], this interpretation suggests that the network might charge users proactively, in hopes of avoiding congestion at links later. From an implementation standpoint, such a shift implies that rather than a hard capacity constraint (i.e., a link is overloaded when the rate through it exceeds the capacity of the link), the link has an elastic capacity (i.e., the link gradually begins to signal a buildup of congestion before the link's true capacity is actually met). Many proposals have been made for "active queue management" (AQM) to achieve good performance with Explicit Congestion Notification; see, e.g., [2, 15, 18, 19]. This issue is of secondary importance to our discussion, as we do not concern ourselves with the specific interpretation of the cost function at the link. (An insightful discussion of the relationship between active queue management and the cost function of the link may be found in [10].)

In this paper, we investigate the robustness of the market mechanism of [16] when users attempt to manipulate the market. Formally, we consider a model where users anticipate the effects of their actions on the link prices. This makes the model a game, and we ask two fundamental questions: first, does a Nash equilibrium exist for this game? And second, how inefficient is such an equilibrium relative to the maximal aggregate surplus? We show that Nash equilibria exist, and that the efficiency loss is no more than a factor $6 - 4\sqrt{2}$ of the maximal aggregate surplus (approximately 34%) when users are price anticipating.

Such an investigation forms part of a broader body of work on quantifying efficiency loss in environments where participants are selfish. Results have been obtained for routing [7, 17, 21], traffic networks [6, 25] and network design problems [1, 8]. Our work is most closely related to that of [12], where the same market mechanism as in this paper was considered for the case where the supply of a link is fixed, or *inelastic*; this was the mechanism first presented in [14]. Johari and Tsitsiklis show the efficiency loss when users are price anticipating is no worse than 25% [12].

The outline of the remainder of the paper is as follows. We start by considering a single link in isolation. In Section 1, we describe the market mechanism for a single link, and recapitulate the results of Kelly et al. [16]. In Section 2, we describe a game where users are price anticipating, and establish the existence of a Nash equilibrium. We also establish necessary and sufficient conditions for a strategy vector to be a Nash equilibrium. These conditions are used in Section 3 to prove the main result of the paper for a single link: that when users are price anticipating, the efficiency loss—that is, the loss in aggregate surplus relative to the maximum—is no more than 34%.

In Section 4, we compare the settings of inelastic and elastic supply. In particular, we consider a limit of cost functions which approach a hard capacity constraint. We show that if these cost functions are monomials and we let the exponent tend to infinity, then the efficiency loss approaches 25%, which is consistent with the result of [12].

In Section 5, we extend the results to general networks. This extension is done using the same approach as [12]. We consider a game where users submit individual bids to each link in the network, and establish existence of a Nash equilibrium. Using techniques similar to the results proven in a network context in [12], we show that the efficiency loss is no more than 34% when users are price anticipating, matching the result of Section 3. Some conclusions are offered in Section 6.

# 1 Background

Suppose $R$ users share a single communication link. Let $d_r \geq 0$ denote the rate allocated to user $r$. We assume that user $r$ receives a *utility* equal to $U_r(d_r)$ if the allocated rate is $d_r$. In addition, we let $f = \sum_r d_r$ denote the total rate allocated at the link, and let $C(f)$ denote the cost incurred at the link when the total allocated rate is $f \geq 0$. We will assume that both $U_r$ and $C$ are measured in the same monetary units. A natural interpretation is that $U_r(d_r)$ is the monetary value to user $r$ of a rate allocation $d_r$, and $C(f)$ is a monetary cost for congestion at the link when the total allocated rate is $f$.

We make the following assumptions regarding $U_r$ and $C$.

**Assumption 1** *For each $r$, over the domain $d_r \geq 0$ the utility function $U_r(d_r)$ is concave, strictly increasing, and continuously differentiable, and the right directional derivative at $0$, denoted $U'_r(0)$, is finite.*

**Assumption 2** *There exists a continuous, convex, strictly increasing function $p(f)$ over $f \geq 0$ with $p(0) = 0$, such that for $f \geq 0$:*

$$C(f) = \int_0^f p(z)dz.$$

*Thus $C(f)$ is strictly convex and strictly increasing.*

Concavity in Assumption 1 corresponds to *elastic* traffic, as defined by Shenker [26]; such traffic includes file transfers such as FTP connections and peer-to-peer connections. Note that Assumption 2 does not require the price function $p$ to be differentiable. Indeed, assuming smoothness of $p$ would simplify some of the technical arguments in the paper. However, we later require the use of nondifferentiable price functions in our proof of Theorem 8.

Given complete knowledge and centralized control of the system, a natural problem for the network manager to try to solve is the following [14]:

*SYSTEM*:

$$\text{maximize} \quad \sum_r U_r(d_r) - C\left(\sum_r d_r\right) \quad (1)$$

$$\text{subject to} \quad d_r \geq 0, \quad r = 1, \ldots, R. \quad (2)$$

Since the objective function (1) is continuous, and $U_r$ increases at most linearly while $C$ increases superlinearly, an optimal solution $\mathbf{d}^S = (d_1^S, \ldots, d_R^S)$ exists for (1)-(2); since the feasible region is convex and $C$ is strictly convex, if the functions $U_r$ are strictly concave, then the optimal solution is unique. We refer to the objective function (1) as the *aggregate surplus*; this is the net monetary benefit to the economy consisting of the users and the single link [20]. For convenience, we define a function $\text{surplus}(\mathbf{d})$ which gives the aggregate surplus at an allocation $\mathbf{d}$:

$$\text{surplus}(\mathbf{d}) \triangleq \sum_r U_r(d_r) - C\left(\sum_r d_r\right). \quad (3)$$

Due to the decentralized nature of the system, the resource manager may not have an exact specification of the utility functions [14]. As a result, we consider the following pricing scheme for rate allocation. Each user $r$ makes a payment (also called a *bid*) of $w_r$ to the resource manager. Given the vector $\mathbf{w} = (w_1, \ldots, w_r)$, the resource manager chooses a rate allocation $\mathbf{d}(\mathbf{w}) = (d_1(\mathbf{w}), \ldots, d_R(\mathbf{w}))$. We assume the manager treats all users alike—in other words, the network manager does not *price differentiate*. Thus the network manager sets a single price $\mu(\mathbf{w})$; we assume that $\mu(\mathbf{w}) = 0$ if $w_r = 0$ for all $r$, and $\mu(\mathbf{w}) > 0$ otherwise. All users are then charged the same price $\mu(\mathbf{w})$, leading to:

$$d_r(\mathbf{w}) = \begin{cases} 0, & \text{if } w_r = 0; \\ \dfrac{w_r}{\mu(\mathbf{w})}, & \text{if } w_r > 0. \end{cases}$$

Associated with this choice of price is an aggregate rate function $f(\mathbf{w})$, defined by:

$$f(\mathbf{w}) = \sum_r d_r(\mathbf{w}) = \begin{cases} 0, & \text{if } \sum_r w_r = 0; \\ \dfrac{\sum_r w_r}{\mu(\mathbf{w})}, & \text{if } \sum_r w_r > 0. \end{cases} \quad (4)$$

We will assume that $w_r$ is measured in the same monetary units as both $U_r$ and $C$. In this case, given a price $\mu > 0$, user $r$ acts to maximize the following payoff function over $w_r \geq 0$:

$$P_r(w_r; \mu) = U_r\left(\frac{w_r}{\mu}\right) - w_r. \quad (5)$$

The first term represents the utility to user $r$ of receiving a rate allocation equal to $w_r/\mu$; the second term is the payment $w_r$ made to the manager. Observe that since utility is measured in monetary units, the payoff is *quasilinear* in money, a typical assumption in modeling market mechanisms [20].

Notice that as formulated above, the payoff function $P_r$ assumes that user $r$ acts as a *price taker*; that is, user $r$ does not *anticipate* the effect of his choice of $w_r$ on the price $\mu$, and hence on his resulting rate allocation $d_r(\mathbf{w})$. Informally, we expect that in such a situation the aggregate surplus will be maximized if the network manager sets a price equal to marginal cost, i.e., if the price function satisfies:

$$\mu(\mathbf{w}) = p(f(\mathbf{w})). \qquad (6)$$

We show in the following proposition that a joint solution to (4) and (6) can be found; we then use this proposition to show that when users optimize (5) and the price is set to satisfy (6), aggregate surplus is maximized.

**Proposition 1** *Suppose Assumption 2 holds. Given any vector of bids $\mathbf{w} \geq 0$, there exists a unique pair $(\mu(\mathbf{w}), f(\mathbf{w})) \geq 0$ satisfying (4) and (6), and in this case $f(\mathbf{w})$ is the unique solution $f$ to:*

$$\sum_r w_r = fp(f). \qquad (7)$$

*Furthermore, $f(\cdot)$ has the following properties: (1) $f(\mathbf{0}) = 0$; (2) $f(\mathbf{w})$ is continuous for $\mathbf{w} \geq 0$; (3) $f(\mathbf{w})$ is a strictly increasing and strictly concave function of $\sum_r w_r$; and (4) $f(\mathbf{w}) \to \infty$ as $\sum_r w_r \to \infty$.*

*Proof.* Fix a vector $\mathbf{w} \geq 0$. First suppose there exists a solution to (4) and (6). Then from (4), we have:

$$\sum_r w_r = f(\mathbf{w})\mu(\mathbf{w}).$$

After substituting (6), the preceding relation becomes (7). Conversely, if $f(\mathbf{w})$ solves (7), then defining $\mu(\mathbf{w})$ according to (6) makes (7) equivalent to (4).

Thus, it suffices to check that there exists a unique solution $f$ to (7). By Assumption 2, $p$ is strictly increasing, and since $p$ is convex, $p(f) \to \infty$ as $f \to \infty$; thus defining $g(f) = fp(f)$, we know that $g(0) = 0$; $g$ is strictly increasing, strictly convex, and continuous; and $g(f) \to \infty$ as $f \to \infty$. Thus $g$ is invertible, and crosses the level $\sum_r w_r$ at a unique value $f(\mathbf{w}) = g^{-1}(\sum_r w_r)$. From this description and the properties of $g$ it is not hard to verify that $f$ has the four properties stated in the proposition. □

Observe that we can view (7) as a market-clearing process. Given the total revenue $\sum_r w_r$ from the users, the link manager chooses an aggregate rate $f(\mathbf{w})$ so that the revenue is exactly equal to the aggregate charge $f(\mathbf{w})p(f(\mathbf{w}))$. Due to Assumption 2, this market-clearing aggregate rate is uniquely determined. Kelly et al. present two algorithms in [16] which amount to dynamic processes of market-clearing; as a result, a key motivation for the mechanism we study in this paper is that it represents the equilibrium behavior of the algorithms in [16]. Kelly et al. show in [16] that when users are non-anticipating, and the network sets the price $\mu(\mathbf{w})$ according to (4) and (6), the resulting allocation solves *SYSTEM*. This is formalized in the following theorem, adapted from [16].

**Theorem 2 (Kelly et al., [16])** *Suppose Assumptions 1 and 2 hold. For any* $\mathbf{w} \geq 0$, *let* $(\mu(\mathbf{w}), f(\mathbf{w}))$ *be the unique solution to* (4) *and* (6). *Then there exists a vector* $\mathbf{w}$ *such that* $\mu(\mathbf{w}) > 0$, *and:*

$$P_r(w_r; \mu(\mathbf{w})) = \max_{\overline{w}_r \geq 0} P_r(\overline{w}_r; \mu(\mathbf{w})), \quad r = 1, \ldots, R. \tag{8}$$

*For any such vector* $\mathbf{w}$, *the vector* $\mathbf{d}(\mathbf{w}) = \mathbf{w}/\mu(\mathbf{w})$ *solves SYSTEM. If the functions* $U_r$ *are strictly concave, such a vector* $\mathbf{w}$ *is unique.*

*Proof.* Let $\mathbf{d}^S$ be any solution to *SYSTEM*; as discussed above, at least one such solution exists. Let $f^S = \sum_r d_r^S$, and define $w_r^S \triangleq d_r^S p(f^S)$ for each $r$. Observe that with this definition, we have $\sum_r w_r^S = \sum_r d_r^S p(f^S) = f^S p(f^S)$; thus $f^S$ satisfies (7), and we have $f(\mathbf{w}^S) = f^S$, $\mathbf{d}(\mathbf{w}^S) = \mathbf{d}^S$.

Given Assumptions 1 and 2, observe that any solution to *SYSTEM* is identified by the following necessary and sufficient optimality conditions:

$$U_r'(d_r^S) = p\left(\sum_s d_s^S\right), \quad \text{if } d_r^S > 0; \tag{9}$$

$$U_r'(0) \leq p\left(\sum_s d_s^S\right), \quad \text{if } d_r^S = 0. \tag{10}$$

Now, since $p(0) = 0$ but $U_r'(0) > 0$ for all $r$, we must have $f^S = \sum_r d_r^S > 0$; thus $\mu(\mathbf{w}) = p(f^S) > 0$. But then $d_r^S = w_r/p(f^S)$ for each $r$, so the preceding optimality conditions become:

$$U_r'\left(\frac{w_r}{p(f^S)}\right) = p(f^S), \quad \text{if } w_r > 0;$$

$$U_r'(0) \leq p(f^S), \quad \text{if } w_r = 0.$$

These conditions ensure that (8) holds.

Conversely, suppose we are given a vector $\mathbf{w}$ such that $\mu(\mathbf{w}) > 0$, and (8) holds. Then we simply reverse the argument above: since (8) holds, we conclude that the optimality conditions (9)-(10) hold with $\mathbf{d}(\mathbf{w}) = \mathbf{w}/\mu(\mathbf{w}) = \mathbf{w}/p(f(\mathbf{w}))$, so that $\mathbf{d}(\mathbf{w})$ is an optimal solution to *SYSTEM*. Finally, if the functions $U_r$ are each strictly concave, then the solution $\mathbf{d}^S$ to *SYSTEM* is unique, so the price $p(f^S)$ is uniquely determined as well. As a result, for each $r$ the product $d_r^S p(f^S)$ is unique, so the vector $\mathbf{w}$ identified in the theorem must be unique as well. □

Theorem 2 shows that with an appropriate choice of price function (as determined by (4) and (6)), and under the assumption that the users behave as price takers, there exists a bid vector $\mathbf{w}$ where all users have optimally chosen their bids $w_r$, with respect to the given price $\mu(\mathbf{w})$; and at this "equilibrium," the aggregate surplus is maximized. However, when the price taking assumption is violated, the model changes into a game and the guarantee of Theorem 2 is no longer valid. We investigate this game in the following section.

## 2 The Single Link Game

We now consider an alternative model where the users of a single link are price anticipating, rather than price taking, and play a game to acquire a share of the link. Throughout the remainder of this section and the next, we will assume that the link manager sets the price $\mu(\mathbf{w})$ according to the unique choice prescribed by Proposition 1, as follows.

**Assumption 3** *For any $\mathbf{w} \geq 0$, the aggregate rate $f(\mathbf{w})$ is the solution to (7): $\sum_r w_r = f(\mathbf{w})p(f(\mathbf{w}))$. Furthermore, for each $r$, $d_r(\mathbf{w})$ is given by:*

$$d_r(\mathbf{w}) = \begin{cases} 0, & \text{if } w_r = 0; \\ \dfrac{w_r}{p(f(\mathbf{w}))}, & \text{if } w_r > 0. \end{cases} \quad (11)$$

Note that we have $f(\mathbf{w}) > 0$ and $p(f(\mathbf{w})) > 0$ if $\sum_r w_r > 0$, and hence $d_r$ is always well defined.

We adopt the notation $\mathbf{w}_{-r}$ to denote the vector of all bids by users other than $r$; i.e., $\mathbf{w}_{-r} = (w_1, w_2, \ldots, w_{r-1}, w_{r+1}, \ldots, w_R)$. Given $\mathbf{w}_{-r}$, each user $r$ chooses $w_r \geq 0$ to maximize:

$$Q_r(w_r; \mathbf{w}_{-r}) \triangleq U_r(d_r(\mathbf{w})) - w_r, \quad (12)$$

over nonnegative $w_r$. The payoff function $Q_r$ is similar to the payoff function $P_r$, except that the user now anticipates that the network will set the price according to Assumption 3, as captured by the allocated rate $d_r(\mathbf{w})$. A *Nash equilibrium* of the game defined by $(Q_1, \ldots, Q_R)$ is a vector $\mathbf{w} \geq 0$ such that for all $r$:

$$Q_r(w_r; \mathbf{w}_{-r}) \geq Q_r(\overline{w}_r; \mathbf{w}_{-r}), \quad \text{for all } \overline{w}_r \geq 0. \quad (13)$$

In the next section, we show that a Nash equilibrium always exists, and give necessary and sufficient conditions for a vector $\mathbf{w}$ to be a Nash equilibrium. In Section 2.2, we outline a class of price functions for which the Nash equilibrium is unique.

### 2.1 Existence of Nash Equilibrium

In this section we establish that a Nash equilibrium exists for the game defined by $(Q_1, \ldots, Q_R)$. We start by establishing certain properties of $d_r(\mathbf{w})$ in the following proposition.

**Proposition 3** *Suppose that Assumptions 1-3 hold. Then: (1) $d_r(\mathbf{w})$ is a continuous function of $\mathbf{w}$; and (2) for any $\mathbf{w}_{-r} \geq 0$, $d_r(\mathbf{w})$ is strictly increasing and concave in $w_r \geq 0$, and $d_r(\mathbf{w}) \to \infty$ as $w_r \to \infty$.*

*Proof.* We first show (1): that $d_r(\mathbf{w})$ is a continuous function of $\mathbf{w}$. Recall from Proposition 1 that $f(\mathbf{w})$ is a continuous function of $\mathbf{w}$, and $f(\mathbf{0}) = 0$. Now at any vector $\mathbf{w}$ such that $\sum_s w_s > 0$, we have $p(f(\mathbf{w})) > 0$, so $d_r(\mathbf{w}) = w_r/p(f(\mathbf{w}))$; thus continuity of $d_r$ at $\mathbf{w}$ follows by continuity of $f$ and $p$. Suppose instead that $\mathbf{w} = \mathbf{0}$, and consider a sequence $\mathbf{w}^n$ such that $\mathbf{w}^n \to \mathbf{0}$ as $n \to \infty$. Then $\sum_r d_r(\mathbf{w}^n) = f(\mathbf{w}^n) \to 0$ as $n \to \infty$, from parts (1) and (2) of Proposition 1; since $d_r(\mathbf{w}^n) \geq 0$ for all $n$, we must have $d_r(\mathbf{w}^n) \to 0 = d_r(\mathbf{0})$ as $n \to \infty$, as required.

We now show (2): that $d_r(\mathbf{w})$ is concave and strictly increasing in $w_r \geq 0$, with $d_r(\mathbf{w}) \to \infty$ as $w_r \to \infty$. From Assumption 3, we can rewrite the definition of $d_r(\mathbf{w})$ as:

$$d_r(\mathbf{w}) = \begin{cases} 0, & \text{if } w_r = 0; \\ \dfrac{w_r}{\sum_s w_s} f(\mathbf{w}), & \text{if } w_r > 0. \end{cases} \quad (14)$$

From this expression and Proposition 1, it follows that $d_r(\mathbf{w})$ is strictly increasing in $w_r$. To show $d_r(\mathbf{w}) \to \infty$ as $w_r \to \infty$, we only need that $f(\mathbf{w}) \to \infty$ as $w_r \to \infty$, a fact that was shown in Proposition 1.

It remains to be shown that for fixed $\mathbf{w}_{-r}$, $d_r$ is a concave function of $w_r \geq 0$. Since we have already shown that $d_r$ is continuous, we may assume without loss of generality that $w_r > 0$. We first assume that $p$ is twice differentiable. In this case, it follows from (7) that $f$ is twice differentiable in $w_r$. Since $w_r > 0$, we can differentiate (14) twice to find:

$$\frac{\partial^2 d_r(\mathbf{w})}{\partial w_r^2} = -\frac{2\sum_{s \neq r} w_s}{(\sum_s w_s)^3} f(\mathbf{w}) + \frac{2\sum_{s \neq r} w_s}{(\sum_s w_s)^2} \cdot \frac{\partial f(\mathbf{w})}{\partial w_r} + \frac{w_r}{\sum_s w_s} \cdot \frac{\partial^2 f(\mathbf{w})}{\partial w_r^2}.$$

From Proposition 1, $f$ is a strictly concave function of $\sum_s w_s$; thus the last term in the sum above is nonpositive. To show that $d_r$ is concave in $w_r$, therefore, it suffices to show that the sum of the first two terms is negative, i.e.:

$$\frac{f(\mathbf{w})}{\sum_s w_s} \geq \frac{\partial f(\mathbf{w})}{\partial w_r}.$$

By differentiating both sides of (7), we find that:

$$\frac{\partial f(\mathbf{w})}{\partial w_r} = \frac{1}{p(f(\mathbf{w})) + f(\mathbf{w})p'(f(\mathbf{w}))}.$$

On the other hand, from (7), we have:

$$\frac{f(\mathbf{w})}{\sum_s w_s} = \frac{1}{p(f(\mathbf{w}))}.$$

Substituting these relations, and noting that $f(\mathbf{w})p'(f(\mathbf{w})) \geq 0$ since $p$ is strictly increasing, we have:

$$\frac{f(\mathbf{w})}{\sum_s w_s} = \frac{1}{p(f(\mathbf{w}))} \geq \frac{1}{p(f(\mathbf{w})) + f(\mathbf{w})p'(f(\mathbf{w}))} = \frac{\partial f(\mathbf{w})}{\partial w_r},$$

as required. Thus $d_r(\mathbf{w})$ is concave in $w_r$, as long as $p$ is twice differentiable.

Now suppose that $p$ is any price function satisfying Assumption 2, but not necessarily twice differentiable. In this case, we may choose a sequence of twice differentiable price functions $p^n$ satisfying Assumption 2, such that $p^n \to p$ pointwise as $n \to \infty$ (i.e., $p^n(f) \to p(f)$ as $n \to \infty$, for all $f \geq 0$).[1] Let $d_r^n$ be the allocation function for user $r$ when the price function is $p^n$; then by the argument in the preceding paragraph, $d_r^n(\mathbf{w})$ is concave in $w_r$, for each $n$. In order to show that $d_r(\mathbf{w})$ is concave in $w_r$, therefore, it suffices to show that $d_r^n \to d_r$ pointwise as $n \to \infty$. From (14), this will be true as long as $f^n \to f$ pointwise as $n \to \infty$, where $f^n$ is the solution to (7) when the price function is $p^n$.

Fix a bid vector $\mathbf{w}$; we now proceed to show that $f^n(\mathbf{w}) \to f(\mathbf{w})$ as $n \to \infty$. For each $n$, define $g^n(f) = fp^n(f)$, and let $g(f) = fp(f)$. By continuity, $g^n(f) \to g(f)$ as $n \to \infty$, for all $f \geq 0$. Furthermore, from (7), $\sum_r w_r = g(f(\mathbf{w}))$. Fix $\varepsilon > 0$, and choose $\delta > 0$ so that:

$$\delta < \min\left\{\sum_r w_r - g(f(\mathbf{w}) - \varepsilon), g(f(\mathbf{w}) + \varepsilon) - \sum_r w_r\right\}.$$

(Note that such a choice is possible because $g$ is strictly increasing.) Now for sufficiently large $n$, we have:

$$g^n(f(\mathbf{w}) - \varepsilon) - g(f(\mathbf{w}) - \varepsilon) < \delta, \quad \text{and } g(f(\mathbf{w}) + \varepsilon) - g^n(f(\mathbf{w}) + \varepsilon) < \delta.$$

From the definition of $\delta$, this yields:

$$g^n(f(\mathbf{w}) - \varepsilon) < \sum_r w_r < g^n(f(\mathbf{w}) + \varepsilon).$$

Since $g^n$ is strictly increasing, and $f^n(\mathbf{w})$ satisfies $g^n(f^n(\mathbf{w})) = \sum_r w_r$, we conclude that $|f^n(\mathbf{w}) - f(\mathbf{w})| < \varepsilon$ for sufficiently large $n$, as required. □

The previous proposition establishes concavity and continuity of $d_r$; this guarantees existence of a Nash equilibrium, as the following proposition shows.

**Proposition 4** *Suppose that Assumptions 1-3 hold. Then there exists a Nash equilibrium $\mathbf{w}$ for the game defined by $(Q_1, \ldots, Q_R)$.*

*Proof.* We begin by observing that we may restrict the strategy space of each user $r$ to a compact set, without loss of generality. To see this, fix a user $r$, and a vector $\mathbf{w}_{-r}$ of bids for all other users. Given a bid $w_r$ for user $r$, we note that $d_r(\mathbf{w}) \leq d_r(w_r; \mathbf{0}_{-r})$, where $\mathbf{0}_{-r}$ denotes the bid vector where all other users bid zero. This inequality follows since $w_r = d_r(\mathbf{w})p(f(\mathbf{w}))$; and if $\sum_s w_s$ decreases, then $p(f(\mathbf{w}))$ decreases as well (from Proposition 1), so $d_r(\mathbf{w})$ must increase.

We thus have $Q_r(w_r; \mathbf{w}_{-r}) \leq U_r(d_r(w_r; \mathbf{0}_{-r})) - w_r$. By concavity of $U_r$, for $w_r > 0$ we have:

$$Q_r(w_r; \mathbf{w}_{-r}) \leq U_r(0) + U_r'(0)d_r(w_r; \mathbf{0}_{-r}) - w_r = U_r(0) + w_r\left(\frac{U_r'(0)}{p(f(w_r; \mathbf{0}_{-r}))} - 1\right). \quad (15)$$

Now observe from (7) that:

$$w_r = f(w_r; \mathbf{0}_{-r})p(f(w_r; \mathbf{0}_{-r})).$$

---

[1] Define $p(f) = 0$ for $f \leq 0$, and consider a sequence of twice differentiable functions $\phi^n$, such that $\phi^n$ has support on $[-1/n, 0]$, and $\int_{-\infty}^{\infty} \phi^n(z)\, dz = 1$. Then it is straightforward to verify the sequence $p^n$ defined by $p^n(f) = \int_{-\infty}^{\infty} p(z)\phi^n(z - f)\, dz$ has the required properties.

Since $p$ is convex and strictly increasing, we have $\lim_{f\to\infty} p(f) = \infty$; thus we conclude that $p(f(w_r; \mathbf{0}_{-r})) \to \infty$ as $w_r \to \infty$. Consequently, using (15), there exists $B_r > 0$ such that if $w_r \geq B_r$, then $Q_r(w_r; \mathbf{w}_{-r}) < U_r(0)$. Since $Q_r(0; \mathbf{w}_{-r}) = U_r(0)$, user $r$ would never choose to bid $w_r \geq B_r$ at a Nash equilibrium. Thus, we may restrict the strategy space of user $r$ to the compact interval $S_r = [0, B_r]$ without loss of generality.

The game defined by $(Q_1, \ldots, Q_R)$ together with the strategy spaces $(S_1, \ldots, S_R)$ is now a *concave R-person game*: applying Proposition 3, each payoff function $Q_r$ is continuous in the composite strategy vector $\mathbf{w}$, and concave in $w_r$ (since $U_r$ is concave and strictly increasing, and $d_r(\mathbf{w})$ is concave in $w_r$); and the strategy space of each user $r$ is a compact, convex, nonempty subset of $\mathbb{R}$. Applying Rosen's existence theorem [24], we conclude that a Nash equilibrium $\mathbf{w}$ exists for this game. $\square$

In the remainder of this section, we establish necessary and sufficient conditions for a vector $\mathbf{w}$ to be a Nash equilibrium. Because the price function $p$ may not be differentiable, we will use *subgradients* to describe necessary local conditions for a vector $\mathbf{w}$ to be a Nash equilibrium. Since the payoff of user $r$ is concave, these necessary conditions will in fact be sufficient for $\mathbf{w}$ to be a Nash equilibrium.

We begin with some concepts from convex analysis [3, 23]. An *extended real-valued function* is a function $g : \mathbb{R} \to [-\infty, \infty]$; such a function is called *proper* if $g(x) > -\infty$ for all $x$, and $g(x) < \infty$ for at least one $x$. We say that a scalar $\gamma$ is a *subgradient* of an extended real-valued function $g$ at $x$ if for all $\overline{x} \in \mathbb{R}$, we have $g(\overline{x}) \geq g(x) + \gamma(\overline{x} - x)$. The *subdifferential* of $g$ at $x$, denoted $\partial g(x)$, is the set of all subgradients of $g$ at $x$. Finally, given an extended real-valued function $g$, we denote the right directional derivative of $g$ at $x$ by $\partial^+ g(x)/\partial x$ and left directional derivative of $g$ at $x$ by $\partial^- g(x)/\partial x$ (if they exist). If $g$ is convex, then $\partial g(x) = [\partial^- g(x)/\partial x, \partial^+ g(x)/\partial x]$, provided the directional derivatives exist.

For the remainder of the paper, we view any price function $p$ as an extended real-valued convex function, by defining $p(f) = \infty$ for $f < 0$. Our first step is a lemma identifying the directional derivatives of $d_r$ as a function of $w_r$; for notational convenience, we introduce the following definitions of $\varepsilon^+(f)$ and $\varepsilon^-(f)$, for $f > 0$:

$$\varepsilon^+(f) \triangleq \frac{f}{p(f)} \cdot \frac{\partial^+ p(f)}{\partial f}, \qquad \varepsilon^-(f) \triangleq \frac{f}{p(f)} \cdot \frac{\partial^- p(f)}{\partial f}. \tag{16}$$

Note that under Assumption 2, we have $0 < \varepsilon^-(f) \leq \varepsilon^+(f)$ for $f > 0$.

**Lemma 5** *Suppose Assumptions 1-3 hold. Then for all $\mathbf{w}$ with $\sum_s w_s > 0$, $d_r(\mathbf{w})$ is directionally differentiable with respect to $w_r$. These directional derivatives are given by:*

$$\frac{\partial^+ d_r(\mathbf{w})}{\partial w_r} = \frac{1}{p(f(\mathbf{w}))} \left(1 - \frac{d_r(\mathbf{w})}{f(\mathbf{w})} \cdot \frac{\varepsilon^+(f(\mathbf{w}))}{1 + \varepsilon^+(f(\mathbf{w}))}\right); \tag{17}$$

$$\frac{\partial^- d_r(\mathbf{w})}{\partial w_r} = \frac{1}{p(f(\mathbf{w}))} \left(1 - \frac{d_r(\mathbf{w})}{f(\mathbf{w})} \cdot \frac{\varepsilon^-(f(\mathbf{w}))}{1 + \varepsilon^-(f(\mathbf{w}))}\right). \tag{18}$$

*Furthermore, $\partial^+ d_r(\mathbf{w})/\partial w_r > 0$, and if $w_r > 0$ then $\partial^- d_r(\mathbf{w})/\partial w_r > 0$.*

*Proof.* Existence of the directional derivatives is obtained because $d_r(\mathbf{w})$ is a concave function of $w_r$ (from Proposition 3). Fix a vector $\mathbf{w}$ of bids, such that $\sum_r w_r > 0$. Since $f$ is an increasing concave function of $w_r$, and the convex function $p$ is directionally differentiable at $f(\mathbf{w})$ ([23], Theorem 23.1), we can apply the chain rule to compute the right directional derivative of (7) with respect to $w_r$:

$$1 = \frac{\partial^+ f(\mathbf{w})}{\partial w_r} \cdot p(f(\mathbf{w})) + f(\mathbf{w}) \cdot \frac{\partial^+ p(f(\mathbf{w}))}{\partial f} \cdot \frac{\partial^+ f(\mathbf{w})}{\partial w_r}.$$

Thus, as long as $\sum_r w_r > 0$, $\partial^+ f(\mathbf{w})/\partial w_r$ exists, and is given by:

$$\frac{\partial^+ f(\mathbf{w})}{\partial w_r} = \left( p(f(\mathbf{w})) + f(\mathbf{w}) \cdot \frac{\partial^+ p(f(\mathbf{w}))}{\partial f} \right)^{-1},$$

We conclude from (11) that the right directional derivative of $d_r(\mathbf{w})$ with respect to $w_r$ is given by:

$$\frac{\partial^+ d_r(\mathbf{w})}{\partial w_r} = \frac{1}{p(f(\mathbf{w}))} - \frac{w_r}{(p(f(\mathbf{w})))^2} \cdot \frac{\partial^+ p(f(\mathbf{w}))}{\partial f} \cdot \frac{\partial^+ f(\mathbf{w})}{\partial w_r}.$$

Simplifying, this reduces to (17). Note that since $d_r(\mathbf{w}) \leq f(\mathbf{w})$ and $\varepsilon^+(f(\mathbf{w}))/[1+\varepsilon^+(f(\mathbf{w}))] < 1$, we have $\partial^+ d_r(\mathbf{w})/\partial w_r > 0$. A similar analysis follows for the left directional derivative. □

For notational convenience, we make the following definitions for $f > 0$:

$$\beta^+(f) \triangleq \frac{\varepsilon^+(f)}{1+\varepsilon^+(f)}, \qquad \beta^-(f) \triangleq \frac{\varepsilon^-(f)}{1+\varepsilon^-(f)}. \tag{19}$$

Under Assumption 2, we have $0 < \beta^-(f) \leq \beta^+(f) < 1$ for $f > 0$.

The next proposition is the central result of this section: it provides simple local conditions that are necessary and sufficient for a vector $\mathbf{w}$ to be a Nash equilibrium.

**Proposition 6** *Suppose that Assumptions 1-3 hold. Then $\mathbf{w}$ is a Nash equilibrium of the game defined by $(Q_1, \ldots, Q_R)$, if and only if $\sum_r w_r > 0$, and with $\mathbf{d} = \mathbf{d}(\mathbf{w})$, $f = f(\mathbf{w})$, the following two conditions hold for all $r$:*

$$U'_r(d_r)\left(1 - \beta^+(f) \cdot \frac{d_r}{f}\right) \leq p(f); \tag{20}$$

$$U'_r(d_r)\left(1 - \beta^-(f) \cdot \frac{d_r}{f}\right) \geq p(f), \quad \text{if } d_r > 0. \tag{21}$$

*Conversely, if $\mathbf{d} \geq 0$ and $f > 0$ satisfy (20)-(21), and $\sum_r d_r = f$, then the vector $\mathbf{w} = p(f)\mathbf{d}$ is a Nash equilibrium with $\mathbf{d} = \mathbf{d}(\mathbf{w})$ and $f = f(\mathbf{w})$.*

*Proof.* We first show that if $\mathbf{w}$ is a Nash equilibrium, then we must have $\sum_r w_r > 0$. Suppose not; then $w_r = 0$ for all $r$. Fix a user $r$; for $\overline{w}_r > 0$, we have $d_r(\overline{w}_r; \mathbf{w}_{-r})/\overline{w}_r = f(\overline{w}_r; \mathbf{w}_{-r})/\overline{w}_r =$ $1/p(f(\overline{w}_r; \mathbf{w}_{-r}))$, which approaches infinity as $\overline{w}_r \to 0$. Thus $\partial^+ d_r(\mathbf{w})/\partial w_r = \infty$, and thus we have:

$$\frac{\partial^+ Q_r(w_r; \mathbf{w}_{-r})}{\partial w_r} = U'_r(0) \cdot \frac{\partial^+ d_r(\mathbf{w})}{\partial w_r} - 1 = \infty.$$

In particular, an infinitesimal increase of $w_r$ strictly increases the payoff of user $r$, so $\mathbf{w} = \mathbf{0}$ cannot be a Nash equilibrium. Thus, if $\mathbf{w}$ is a Nash equilibrium, then $\sum_r w_r > 0$.

Now let $\mathbf{w}$ be a Nash equilibrium. We established in Lemma 5 that $d_r$ is directionally differentiable in $w_r$ for each $r$, as long as $\sum_s w_s > 0$. Thus, from (13), if $\mathbf{w}$ is a Nash equilibrium, then the following two conditions must hold:

$$\frac{\partial^+ Q_r(w_r; \mathbf{w}_{-r})}{\partial w_r} = U'_r(d_r(\mathbf{w})) \cdot \frac{\partial^+ d_r(\mathbf{w})}{\partial w_r} - 1 \leq 0;$$

$$\frac{\partial^- Q_r(w_r; \mathbf{w}_{-r})}{\partial w_r} = U'_r(d_r(\mathbf{w})) \cdot \frac{\partial^- d_r(\mathbf{w})}{\partial w_r} - 1 \geq 0, \quad \text{if } w_r > 0.$$

We may substitute using Lemma 5 to find that if $\mathbf{w}$ is a Nash equilibrium, then:

$$U'_r(d_r(\mathbf{w})) \left(1 - \beta^+(f(\mathbf{w})) \cdot \frac{d_r(\mathbf{w})}{f(\mathbf{w})}\right) \leq p(f(\mathbf{w}));$$

$$U'_r(d_r(\mathbf{w})) \left(1 - \beta^-(f(\mathbf{w})) \cdot \frac{d_r(\mathbf{w})}{f(\mathbf{w})}\right) \geq p(f(\mathbf{w})), \quad \text{if } w_r > 0.$$

Since the condition $w_r > 0$ is identical to the condition $d_r(\mathbf{w}) > 0$, this establishes the conditions in the proposition. Conversely, if $\sum_r w_r > 0$ and the preceding two conditions hold, then we may reverse the argument: since the payoff function of user $r$ is a concave function of $w_r$ for each $r$ (from Proposition 3), (20)-(21) are sufficient for $\mathbf{w}$ to be a Nash equilibrium.

Finally, suppose that $\mathbf{d}$ and $f > 0$ satisfy (20)-(21), with $\sum_r d_r = f$. Then let $w_r = d_r p(f)$. We then have $\sum_r w_r > 0$ (since $f > 0$); and $\sum_r w_r = fp(f)$, so that $f = f(\mathbf{w})$. Finally, since $f > 0$, we have $d_r = w_r/p(f) = w_r/p(f(\mathbf{w}))$, so that $d_r = d_r(\mathbf{w})$. Thus $\mathbf{w}$ is a Nash equilibrium, as required. □

Note that the preceding proposition identifies a Nash equilibrium entirely in terms of the allocation made; and conversely, if we find a pair $(\mathbf{d}, f)$ which satisfies (20)-(21) with $f > 0$ and $\sum_r d_r = f$, then there exists a Nash equilibrium which yields that allocation. In particular, the set of allocations $\mathbf{d}$ which can arise at Nash equilibria coincides with those vectors $\mathbf{d}$ such that $f = \sum_r d_r > 0$, and (20)-(21) are satisfied.

## 2.2 Nondecreasing Elasticity Price Functions: Uniqueness of Nash Equilibrium

In this section, we demonstrate that for a certain class of differentiable price functions, there exists a *unique* Nash equilibrium of the game defined by $(Q_1, \ldots, Q_R)$. We consider price functions $p$ which satisfy the following additional assumption.

**Assumption 4** *The price function $p$ is differentiable, and exhibits* nondecreasing elasticity*: for $0 < f_1 \leq f_2$, there holds:*
$$\frac{f_1 p'(f_1)}{p(f_1)} \leq \frac{f_2 p'(f_2)}{p(f_2)}.$$

To gain some intuition for the concept of nondecreasing elasticity, consider a price function $p$ satisfying Assumption 2. The quantity $fp'(f)/p(f)$ is known as the *elasticity* of a price function $p$ [20]. Note that the elasticity of $p(f)$ is the derivative of $\ln(p(f))$ with respect to $\ln f$. From this viewpoint, we see that nondecreasing elasticity is equivalent to the requirement that $\ln(p(f))$ is a convex function in $\ln f$. (Note that this is not equivalent to the requirement that $p$ is a convex function of $f$.)

Nondecreasing elasticity can also be interpreted by considering the price function as the inverse of the *supply function* $s(\mu) = p^{-1}(\mu)$; the supply function gives the amount of rate the provider is willing to supply at a given price $\mu$ [20]. In this case, nondecreasing elasticity of the price function is equivalent to nonincreasing elasticity of the supply function.

Nondecreasing elasticity captures a wide range of price functions; we give two common examples below.

**Example 1 (The M/M/1 Queue)** Consider the cost function $C(f) = af/(s - f)$, where $a > 0$ and $s > 0$ are constants; then the cost is proportional to the steady-state queue size in an M/M/1 queue with service rate $s$ and arrival rate $f$. (Note that we must view $p$ as an extended real-valued function, with $p(f) = \infty$ for $f > s$; this does not affect any of the analysis of this paper.) It is straightforward to check that, as long as $0 < f < s$, we have:
$$\frac{fp'(f)}{p(f)} = \frac{2f}{s - f},$$
which is a strictly increasing function of $f$. Thus $p$ satisfies Assumption 4.

**Example 2 (M/M/1 Overflow Probability)** Consider the function $p(f) = a(f/s)^B$, where $a > 0$, $s > 0$, and $B \geq 1$ is an integer. Then the price is set proportional to the probability that an M/M/1 queue exceeds a buffer level $B$, when the service rate is $s$ and the arrival rate is $f$. In this case we have $fp'(f)/p(f) = B$, so that $p$ satisfies Assumption 4.

We now prove the key property of differentiable nondecreasing elasticity price functions in the current development: for such functions, there exists a unique Nash equilibrium of the game defined by $(Q_1, \ldots, Q_R)$.

**Proposition 7** *Suppose Assumptions 1-3 hold. If in addition $p$ is differentiable and exhibits nondecreasing elasticity (Assumption 4 holds), then there exists a unique Nash equilibrium for the game defined by $(Q_1, \ldots, Q_R)$.*

*Proof.* We use the expressions (20)-(21) to show that the Nash equilibrium is unique under Assumption 4. Observe that in this case, from (19), we may define $\beta(f) = \beta^+(f) = \beta^-(f)$ for $f > 0$, and conclude that $\mathbf{w}$ is a Nash equilibrium if and only if $\sum_s w_s > 0$ and the following optimality conditions hold:

$$U_r'(d_r(\mathbf{w})) \left(1 - \beta(f(\mathbf{w})) \cdot \frac{d_r(\mathbf{w})}{f(\mathbf{w})}\right) = p(f(\mathbf{w})), \quad \text{if } w_r > 0; \qquad (22)$$

$$U_r'(0) \leq p(f(\mathbf{w})), \quad \text{if } w_r = 0. \qquad (23)$$

Suppose we have two Nash equilibria $\mathbf{w}^1$, $\mathbf{w}^2$, with $0 < \sum_s w_s^1 < \sum_s w_s^2$; then $p(f(\mathbf{w}^1)) < p(f(\mathbf{w}^2))$, and $f(\mathbf{w}^1) < f(\mathbf{w}^2)$. Note that $U_r'(d_r)$ is nonincreasing as $d_r$ increases; and $\beta(f)$ is nondecreasing as $f$ increases (from Assumption 4), and therefore $\beta(f(\mathbf{w}^1)) \leq \beta(f(\mathbf{w}^2))$. Furthermore, if $w_r^2 > 0$, then from (22) we have $U_r'(0) > p(f(\mathbf{w}^2))$; thus $U_r'(0) > p(f(\mathbf{w}^1))$, so $w_r^1 > 0$ as well (from (23)).

Now note that the right hand side of (22) is strictly larger at $\mathbf{w}^1$ than at $\mathbf{w}^2$; thus the left hand side must be strictly larger at $\mathbf{w}^1$ than at $\mathbf{w}^2$ as well. This is only possible if $d_r(\mathbf{w}^1)/f(\mathbf{w}^1) > d_r(\mathbf{w}^2)/f(\mathbf{w}^2)$ for each user $r$, since we have shown in the preceding paragraph that $f(\mathbf{w}^1) < f(\mathbf{w}^2)$; $U_r'(d_r)$ is nonincreasing as $d_r$ increases; and $\beta(f(\mathbf{w}^1)) \leq \beta(f(\mathbf{w}^2))$. Since $f(\mathbf{w}) = \sum_r d_r(\mathbf{w})$, we have:

$$1 = \sum_{r: w_r^2 > 0} \frac{d_r(\mathbf{w}^2)}{f(\mathbf{w}^2)} < \sum_{r: w_r^2 > 0} \frac{d_r(\mathbf{w}^1)}{f(\mathbf{w}^1)} = 1,$$

which is a contradiction. Thus at the two Nash equilibria, we must have $\sum_s w_s^1 = \sum_s w_s^2$, so we can let $f_0 = f(\mathbf{w}^1) = f(\mathbf{w}^2)$, $p_0 = p(f_0)$, and $\beta_0 = \beta(f_0)$. Then all Nash equilibria $\mathbf{w}$ satisfy:

$$U_r'(d_r(\mathbf{w})) \left(1 - \beta_0 \frac{d_r(\mathbf{w})}{f_0}\right) = p_0, \quad \text{if } w_r > 0; \qquad (24)$$

$$U_r'(0) \leq p_0, \quad \text{if } w_r = 0. \qquad (25)$$

But now we observe that the left hand side of (24) is strictly decreasing in $d_r(\mathbf{w})$, so given $p_0$, there exists at most one solution $d_r(\mathbf{w})$ to (24). Since $w_r = d_r(\mathbf{w})p_0$, this implies the Nash equilibrium $\mathbf{w}$ must be unique. □

We observe that uniqueness of the Nash equilibrium implies an additional desirable property in the case of symmetric users. If two users share the same utility function, and the price function $p$ is differentiable, we conclude from Proposition 7 that at the unique Nash equilibrium, these users submit exactly the same bid (and hence receive exactly the same rate allocation).

## 3 Efficiency Loss: The Single Link Case

We let $\mathbf{d}^S$ denote an optimal solution to *SYSTEM*, defined in (1)-(2), and let $\mathbf{w}$ denote any Nash equilibrium of the game defined by $(Q_1, \ldots, Q_R)$. We now investigate the efficiency loss of this system; that is, how much aggregate surplus is lost because the users attempt to "game" the system? To answer this question, we must compare the aggregate surplus $\sum_r U_r(d_r(\mathbf{w})) - C(\sum_r d_r(\mathbf{w}))$ obtained when the users fully evaluate the effect of their actions on the price, and the aggregate surplus $\sum_r U_r(d_r^S) - C(\sum_r d_r^S)$ obtained by choosing an allocation which maximizes aggregate surplus. The following theorem is the main result of this paper: it states that the efficiency loss is no more than approximately 34%, and that this bound is essentially tight.

**Theorem 8** *Suppose that Assumptions 1-3 hold. Suppose also that $U_r(0) \geq 0$ for all $r$. Let $\mathbf{d}^S$ be any solution to SYSTEM, and let $\mathbf{w}$ be any Nash equilibrium of the game defined by $(Q_1, \ldots, Q_R)$. Then we have the following bound:*

$$\text{surplus}(\mathbf{d}(\mathbf{w})) \geq \left(4\sqrt{2} - 5\right) \cdot \text{surplus}(\mathbf{d}^S), \tag{26}$$

*where* $\text{surplus}(\cdot)$ *is defined in* (3). *In other words, there is no more than approximately a 34% efficiency loss when users are price anticipating.*

*Furthermore, this bound is tight: for every $\delta > 0$, there exists a choice of $R$, a choice of (linear) utility functions $U_r$, $r = 1, \ldots, R$, and a (piecewise linear) price function $p$ such that a Nash equilibrium $\mathbf{w}$ and a solution $\mathbf{d}^S$ to SYSTEM exist with:*

$$\text{surplus}(\mathbf{d}(\mathbf{w})) \leq \left(4\sqrt{2} - 5 + \delta\right) \cdot \text{surplus}(\mathbf{d}^S). \tag{27}$$

*Proof.* The proof of (26) consists of a sequence of steps:

1. We show that the worst case ratio occurs when the utility function of each user is linear.

2. We restrict attention to games where the total allocated Nash equilibrium rate is $f = 1$.

3. We compute the worst case choice of linear utility functions, for a fixed price function $p(\cdot)$ and total Nash equilibrium rate $f = 1$.

4. We prove that it suffices to consider a special class of piecewise linear price functions.

5. Combining Steps 1-3, we compute the worst case efficiency loss by minimizing the ratio of Nash equilibrium aggregate surplus to maximal aggregate surplus, over the worst case choice of games with linear utility functions (from Step 2) and our restricted class of piecewise linear price functions (from Step 3).

*Step 1: Show that we may assume without loss of generality that $U_r$ is linear for each user $r$; i.e., without loss of generality we may assume $U_r(d_r) = \alpha_r d_r$, where $\alpha_1 = 1$ and $0 < \alpha_r \leq 1$ for $r > 1$.* The proof of this claim is similar to the proof of Lemma 4 in [12]. Let $\mathbf{d}^S$ denote any solution to *SYSTEM*, and let $\mathbf{w}$ denote a Nash equilibrium, for an arbitrary collection of utility functions $(U_1, \ldots, U_R)$ satisfying the assumptions of the theorem. We let $\mathbf{d} = \mathbf{d}(\mathbf{w})$ denote the allocation vector at the Nash equilibrium. For each user $r$, we define a new utility function $\overline{U}_r(d_r) = \alpha_r d_r$, where $\alpha_r = U_r'(d_r)$; we know that $\alpha_r > 0$ by Assumption 1. Then observe that if we replace the utility functions $(U_1, \ldots, U_R)$ with the linear utility functions $(\overline{U}_1, \ldots, \overline{U}_R)$, the vector $\mathbf{w}$ remains a Nash equilibrium; this follows from the necessary and sufficient conditions of Proposition 6.

We first show that $\sum_r \alpha_r d_r - C(f) > 0$. To see this, note from (21) that $\alpha_r > p(f)$ for all $r$ such that $d_r > 0$. Thus $\alpha_r d_r > d_r p(f)$ for such a user $r$, so $\sum_r \alpha_r d_r > f p(f) \geq C(f)$, by convexity (Assumption 2).

Next, we note that $\sum_r U_r(d_r^S) - C(\sum_r d_r^S) > 0$. This follows since $U_r$ is strictly increasing and nonnegative, while $C'(0) = p(0) = 0$; thus if $\overline{d}_r$ is sufficiently small for all $r$, we will have $\sum_r U_r(\overline{d}_r) - C(\sum_r \overline{d}_r) > 0$, which implies $\sum_r U_r(d_r^S) - C(\sum_r d_r^S) > 0$ (since $\mathbf{d}^S$ is a solution to *SYSTEM*).

Using concavity, we have for each $r$:

$$U_r(d_r^S) \leq U_r(d_r) + \alpha_r(d_r^S - d_r).$$

Expanding the definition of surplus$(\cdot)$, we have:

$$\frac{\text{surplus}(\mathbf{d})}{\text{surplus}(\mathbf{d}^S)} = \frac{\sum_r U_r(d_r) - C(\sum_r d_r)}{\sum_r U_r(d_r^S) - C(\sum_r d_r^S)}$$

$$\geq \frac{\sum_r \left(U_r(d_r) - \alpha_r d_r\right) + \sum_r \alpha_r d_r - C(\sum_r d_r)}{\sum_r \left(U_r(d_r) - \alpha_r d_r\right) + \sum_r \alpha_r d_r^S - C(\sum_r d_r^S)}$$

$$\geq \frac{\sum_r \left(U_r(d_r) - \alpha_r d_r\right) + \sum_r \alpha_r d_r - C(\sum_r d_r)}{\sum_r \left(U_r(d_r) - \alpha_r d_r\right) + \max_{\overline{\mathbf{d}} \geq 0} \left(\sum_r \alpha_r \overline{d}_r - C(\sum_r \overline{d}_r)\right)}.$$

(Note that all denominators are positive, since we have shown that $\sum_r U_r(d_r^S) - C(\sum_r d_r^S) > 0$.) Since we assumed $U_r(0) \geq 0$, we have $U_r(d_r) - U_r'(d_r)d_r \geq 0$ by concavity; and since $0 < \sum_r \alpha_r d_r - C(f) \leq \max_{\overline{\mathbf{d}} \geq 0}(\sum_r \alpha_r \overline{d}_r - C(\sum_r \overline{d}_r))$, we have the inequality:

$$\frac{\sum_r U_r(d_r) - C(\sum_r d_r)}{\sum_r U_r(d_r^S) - C(\sum_r d_r^S)} \geq \frac{\sum_r \alpha_r d_r - C(\sum_r d_r)}{\max_{\overline{\mathbf{d}} \geq 0} \left(\sum_r \alpha_r \overline{d}_r - C(\sum_r \overline{d}_r)\right)}.$$

Now observe that the right hand side of the previous expression is the ratio of the Nash equilibrium aggregate surplus to the maximal aggregate surplus, when the utility functions are $(\overline{U}_1, \ldots, \overline{U}_R)$; since this ratio is no larger than the same ratio for the original utility functions $(U_1, \ldots, U_R)$, we can restrict attention to games where the utility function of each user is linear. Finally, by replacing $\alpha_r$ by $\alpha_r/(\max_s \alpha_s)$, and the cost function $C(\cdot)$ by $C(\cdot)/(\max_s \alpha_s)$, we may assume without loss of generality that $\max_r \alpha_r = 1$. Thus, by relabeling the users if necessary, we assume for the remainder of the proof that $U_r(d_r) = \alpha_r d_r$ for all $r$, where $\alpha_1 = 1$ and $0 < \alpha_r \leq 1$ for $r > 1$.

Before continuing, we observe that under these conditions, we have the following relation:

$$\max_{\overline{\mathbf{d}} \geq 0} \left(\sum_r \alpha_r \overline{d}_r - C\left(\sum_r \overline{d}_r\right)\right) = \max_{\overline{f} \geq 0} \left(\overline{f} - C(\overline{f})\right).$$

To see this, note that at any fixed value of $\overline{f} = \sum_r \overline{d}_r$, the left hand side is maximized by allocating the entire rate $\overline{f}$ to user 1. Thus, the ratio of Nash equilibrium aggregate surplus to maximal aggregate surplus becomes:

$$\frac{\sum_r \alpha_r d_r - C(\sum_r d_r)}{\max_{\overline{f} \geq 0} \left( \overline{f} - C(\overline{f}) \right)}. \tag{28}$$

Note that the denominator is positive, since $C'(0) = p(0) = 0$; and further, the optimal solution in the denominator occurs at the unique value of $\overline{f} > 0$ such that $p(\overline{f}) = 1$.

*Step 2: Show that we may restrict attention to games where the total allocated rate at the Nash equilibrium is $f = 1$.* Fix a cost function $C$ satisfying Assumption 2. Let $\mathbf{w}$ be a Nash equilibrium, and let $\mathbf{d} = \mathbf{d}(\mathbf{w})$ be the resulting allocation. Let $f = \sum_r d_r$ be the total allocated rate at the Nash equilibrium; note that $f > 0$ by Proposition 6. We now define a new price function $\hat{p}$ according to $\hat{p}(\hat{f}) = p(f \cdot \hat{f})$, and a new cost function $\hat{C}(\hat{f}) = \int_0^{\hat{f}} \hat{p}(z)\, dz$; note that $\hat{C}(\hat{f}) = C(f \cdot \hat{f})/f$. Then it is straightforward to check that $\hat{p}$ satisfies Assumption 2. We will use hats to denote the corresponding functions when the price function is $\hat{p}$: $\hat{\beta}^+(\hat{f})$, $\hat{\beta}^-(\hat{f})$, $\hat{d}_r(\mathbf{w})$, $\hat{f}(\mathbf{w})$, etc.

Define $\hat{w}_r = w_r/f$. Then we claim that $\hat{\mathbf{w}}$ is a Nash equilibrium when the price function is $\hat{p}$. First observe that:

$$\sum_r \hat{w}_r = \frac{\sum_r w_r}{f} = p(f) = \hat{p}(1).$$

Thus $\hat{f}(\hat{\mathbf{w}}) = 1$. Furthermore:

$$\hat{d}_r(\hat{\mathbf{w}}) = \frac{\hat{w}_r}{\hat{p}(\hat{f}(\hat{\mathbf{w}}))} = \frac{\hat{w}_r}{\hat{p}(1)} = \frac{w_r}{fp(f)} = \frac{d_r}{f}.$$

Finally, note that:

$$\frac{\partial^+ \hat{p}(1)}{\partial \hat{f}} = f \frac{\partial^+ p(f)}{\partial f},$$

from which we conclude that $\hat{\beta}^+(1) = \beta^+(f)$, and similarly $\hat{\beta}^-(1) = \beta^-(f)$. Recall that $\mathbf{w}$ is a Nash equilibrium for the price function $p$; thus, if we combine the preceding conclusions and apply Proposition 6, we have that $\hat{\mathbf{w}}$ is a Nash equilibrium when the price function is $\hat{p}$, with total allocated rate $\hat{f} = 1$ and allocation $\hat{\mathbf{d}} = \mathbf{d}/f$.

To complete the proof of this step, we note the following chain of inequalities:

$$\frac{\sum_r \alpha_r d_r - C(\sum_r d_r)}{\max_{\hat{f} \geq 0} \left( \hat{f} - C(\hat{f}) \right)} = \frac{\sum_r \alpha_r \hat{d}_r - \hat{C}(1)}{\max_{\hat{f} \geq 0} \left( \hat{f}/f - C(\hat{f})/f \right)} \tag{29}$$

$$= \frac{\sum_r \alpha_r \hat{d}_r - \hat{C}(1)}{\max_{g \geq 0} \left( g - \hat{C}(g) \right)}, \tag{30}$$

where we make the substitution $g = \hat{f}/f$. But now note that the right hand side is the ratio of Nash equilibrium aggregate surplus to maximal aggregate surplus for a game where the total allocated rate at the Nash equilibrium is equal to 1. Consequently, in computing the worst case efficiency loss, we may restrict our attention to games where the Nash equilibrium allocated rate is equal to 1.

*Step 3: For a fixed price function $p$, determine the instance of linear utility functions that minimizes Nash equilibrium aggregate surplus, for a fixed Nash equilibrium allocated rate $f = \sum_r d_r = 1$. Note that fixing the price function $p$ fixes the optimal aggregate surplus; thus minimizing the aggregate surplus at Nash equilibrium also yields the worst case efficiency loss.*

We will optimize over the set of all games where users have linear utility functions (satisfying the conditions of Step 1), and where the total Nash equilibrium rate is $f = 1$. We use the necessary and sufficient conditions of Proposition 6. Note that by fixing the price function $p$ and the total rate $f > 0$, the Nash equilibrium price is fixed, $p(1)$, and $\beta^+(1)$ and $\beta^-(1)$ are fixed as well (from the definition (19)); for notational convenience, we abbreviate $p = p(1)$, $C = C(1)$, $\beta^+ = \beta^+(1)$, and $\beta^- = \beta^-(1)$ for the duration of this step. Since $\alpha_1 = 1$, for a fixed value of $R$ the game with linear utility functions that minimizes aggregate surplus is given by solving the following optimization problem (with unknowns $d_1, \ldots, d_R, \alpha_2, \ldots, \alpha_R$):

$$\text{minimize} \quad d_1 + \sum_{r=2}^{R} \alpha_r d_r - C \tag{31}$$

$$\text{subject to} \quad \alpha_r \left(1 - \beta^+ d_r\right) \leq p, \quad r = 1, \ldots, R; \tag{32}$$

$$\alpha_r \left(1 - \beta^- d_r\right) \geq p, \quad \text{if } d_r > 0, \ r = 1, \ldots, R; \tag{33}$$

$$\sum_{r=1}^{R} d_r = 1; \tag{34}$$

$$0 < \alpha_r \leq 1, \quad r = 2, \ldots, R; \tag{35}$$

$$d_r \geq 0, \quad r = 1, \ldots, R. \tag{36}$$

(Note that we have applied Proposition 6: if we solve the preceding problem and find an allocation $\mathbf{d}$ and coefficients $\boldsymbol{\alpha}$, then there exists a Nash equilibrium $\mathbf{w}$ with $\mathbf{d} = \mathbf{d}(\mathbf{w})$.) The objective function is the aggregate surplus given a Nash equilibrium allocation $\mathbf{d}$. The conditions (32)-(33) are equivalent to the Nash equilibrium conditions established in Proposition 6. The constraint (34) ensures that the total allocation made is equal to 1, and the constraint (35) follows from Step 1. The constraint (36) ensures the rate allocated to each user is nonnegative.

We solve this problem through a sequence of reductions. We first show we may assume without loss of generality that the constraint (33) holds with equality for all users $r = 2, \ldots, R$. The resulting problem is symmetric in the users $r = 2, \ldots, R$; we next show that a feasible solution exists if and only if $1 - \beta^+ \leq p < 1$ and $R$ is sufficiently large, and we conclude using a convexity argument that $d_r = (f - d_1)/(R - 1)$ at an optimal solution. Finally, we show the worst case occurs in the limit where $R \to \infty$, and calculate the resulting Nash equilibrium aggregate surplus.

We first show that it suffices to optimize over all $(\boldsymbol{\alpha}, \mathbf{d})$ such that (33) holds with equality for $r = 2, \ldots, R$. Note that if $(\boldsymbol{\alpha}, \mathbf{d})$ is a feasible solution to (31)-(36), then from (33)-(36), and the fact that $0 < \beta^- < 1$, we conclude that $p < 1$. Now if $d_r > 0$ for some $r = 2, \ldots, R$, but the corresponding constraint in (33) does not hold with equality, we can reduce $\alpha_r$ until the constraint in (33) does hold with equality; by this process we obtain a smaller value for the objective function (31). On the other hand, if $d_r = 0$ for some $r = 2, \ldots, R$, we can set $\alpha_r = p$; since $p < 1$, this preserves feasibility, but does not impact the term $\alpha_r d_r$ in the objective function (31). We can therefore restrict attention to feasible solutions for which:

$$\alpha_r = \frac{p}{1 - \beta^- d_r}, \quad r = 2, \ldots, R. \tag{37}$$

Having done so, observe that the constraint (35) that $\alpha_r \leq 1$ may be written as:

$$d_r \leq \frac{1 - p}{\beta^-}, \quad r = 2, \ldots, R.$$

Finally, the constraint (35) that $\alpha_r > 0$ becomes redundant, as it is guaranteed by the fact that $d_r \leq 1$ (from (34)), $\beta^- < 1$ (by definition), and (37).

We now use the preceding observations to simplify the optimization problem (31)-(36) as follows:

$$\text{minimize} \quad d_1 + p \sum_{r=2}^{R} \frac{d_r}{1 - \beta^- d_r} - C \tag{38}$$

$$\text{subject to} \quad 1 - \beta^+ d_1 \leq p \leq 1 - \beta^- d_1; \tag{39}$$

$$\sum_{r=1}^{R} d_r = 1; \tag{40}$$

$$d_r \leq \frac{1 - p}{\beta^-}, \quad r = 2, \ldots, R; \tag{41}$$

$$d_r \geq 0, \quad r = 1, \ldots, R. \tag{42}$$

The objective function (38) equals (31) upon substitution for $\alpha_r$ for $r = 2, \ldots, R$, from (37). We know that $d_1 > 0$ when $p(f) < 1$ (from (32)-(33)); thus the constraint (39) is equivalent to the constraints (32)-(33) for user 1 with $d_1 > 0$. The constraint (32) for $r > 1$ is redundant and eliminated, since (33) holds with equality for $r > 1$. The constraint (40) is equivalent to the allocation constraint (34); and the constraint (41) ensures $\alpha_r \leq 1$, as required in (35).

We first note that for a feasible solution to (38)-(42) to exist, we must have $1 - \beta^+ \leq p < 1$. We have already shown that we must have $p < 1$ if a feasible solution exists. Furthermore, from (39) we observe that the smallest feasible value of $d_1$ is $d_1 = (1 - p)/\beta^+$. We require $d_1 \leq 1$ from (40) and (42), so we must have $(1 - p)/\beta^+ \leq 1$, which yields the restriction that $1 - \beta^+ \leq p$. Thus, there only exist Nash equilibria with total rate 1 and price $p$ if:

$$1 - \beta^+ \leq p < 1. \tag{43}$$

We will assume for the remainder of this step that (43) is satisfied.

We note that if $\overline{\mathbf{d}} = (\overline{d}_1, \ldots, \overline{d}_R)$ is a feasible solution to (38)-(42) with $R$ users, then letting $\overline{d}_{R+1} = 0$, the vector $(\overline{d}_1, \ldots, \overline{d}_{R+1})$ is a feasible solution to (38)-(42) with $R + 1$ users, and with the same objective function value (38) as $\overline{\mathbf{d}}$. Thus, the minimal objective function value cannot increase as $R$ increases, so the worst case efficiency loss occurs in the limit where $R \to \infty$.

We now solve (38)-(42) for a fixed feasible value of $d_1$. From the constraints (40)-(41), we observe that a feasible solution to (38)-(42) exists if and only if the following condition holds in addition to (43):

$$d_1 + (R-1) \cdot \frac{1-p}{\beta^-} \geq 1. \tag{44}$$

In this case, the following symmetric solution is feasible:

$$d_r = \frac{1-d_1}{R-1}, \quad r = 2, \ldots, R. \tag{45}$$

Furthermore, since the objective function is strictly convex and symmetric in the variables $d_2, \ldots, d_R$, and the feasible region is convex, the symmetric solution (45) must be optimal.

If we substitute the optimal solution (45) into the objective function (38) and take the limit as $R \to \infty$, then the constraint (44) is vacuously satisfied, and the objective function becomes $d_1 + p(1-d_1) - C$. Since we have shown that $p < 1$, the worst case occurs at the smallest feasible value of $d_1$; from (39), this value is:

$$d_1 = \frac{1-p}{\beta^+}. \tag{46}$$

The resulting worst case Nash equilibrium aggregate surplus is:

$$p + \frac{(1-p)^2}{\beta^+} - C.$$

To complete the proof of the theorem, we will consider the ratio of this Nash equilibrium aggregate surplus to the maximal aggregate surplus; we denote this ratio by $F(p)$ as a function of the price function $p(\cdot)$:

$$F(p) = \frac{p(1) + (1-p(1))^2/\beta^+(1) - C(1)}{\max_{f \geq 0} (f - C(f))}. \tag{47}$$

Note that henceforth, the scalar $p$ used throughout Step 3 will be denoted $p(1)$, and we return to denoting the price function by $p$. Thus $F(p)$ as defined in (47) is a function of the entire price function $p(\cdot)$.

For completeness, we summarize in the following lemma an intermediate tightness result which will be necessary to prove the tightness of the bound in the theorem.

**Lemma 9** *Suppose that Assumptions 2 and 3 are satisfied. Then there exists $R > 0$ and a choice of linear utility functions $U_r(d_r) = \alpha_r d_r$, where $\alpha_1 = \max_s \alpha_s = 1$, with total Nash equilibrium rate 1, if and only if (43) is satisfied, i.e.:*

$$1 - \beta^+(1) \leq p(1) < 1. \tag{48}$$

*In this case, given $\delta > 0$, there exists $R > 0$ and a collection of $R$ users where user $r$ has utility function $U_r(d_r) = \alpha_r d_r$, such that $\mathbf{d}$ is a Nash equilibrium allocation with $\sum_r d_r = 1$, and:*

$$\frac{\sum_r \alpha_r d_r - C(1)}{\max_{\mathbf{d} \geq 0} (\sum_r \alpha_r d_r - C(\sum_r d_r))} \leq F(p) + \delta. \tag{49}$$

*Proof of Lemma.* The proof follows from Step 3. We have shown that if there exists a Nash equilibrium with total rate 1, then (48) must be satisfied. Conversely, if (48) is satisfied, we proceed as follows: define $d_1$ according to (46); choose $R$ large enough that (44) is satisfied; define $d_r$ according to (45); and then define $\alpha_r$ according to (37) with $p = p(1)$. Then it follows that $(\mathbf{d}, \boldsymbol{\alpha})$ is a feasible solution to (31)-(36), which (by Proposition 6) guarantees there exists a Nash equilibrium whose total allocated rate equals 1.

The bound in (49) then follows by the proof of Step 3. □

The remainder of the proof amounts to minimizing the worst case ratio of Nash equilibrium aggregate surplus to maximal aggregate surplus, over all valid choices of $p$. A valid choice of $p$ is any price function $p$ such that at least one choice of linear utility functions satisfying the conditions of Step 1 leads to a Nash equilibrium with total allocated rate 1. By Lemma 9, all such functions $p$ are characterized by the constraint (48). We will minimize $F(p)$, given by (47), over all choices of $p$ satisfying (48).

*Step 4: Show that in minimizing $F(p)$ over $p$ satisfying* (48), *we may restrict attention to functions $p$ satisfying the following conditions:*

$$p(f) = \begin{cases} af, & 0 \leq f \leq 1; \\ a + b(f-1), & f \geq 1; \end{cases} \quad (50)$$

$$0 < a \leq b; \quad (51)$$

$$\frac{1}{a+b} \leq 1 < \frac{1}{a}. \quad (52)$$

Observe that $p$ as defined in (50)-(52) is a convex, strictly increasing, piecewise linear function with two parts: an initial segment which increases at slope $a > 0$, and a second segment which increases at slope $b \geq a$. In particular, such a function satisfies Assumption 2. Furthermore, we have $\partial^+ p(1)/\partial f = b$, so that $\varepsilon^+(1) = b/a$. This implies $\beta^+(1) = b/(a+b)$; thus, multiplying through (52) by $a$ yields (48).

To verify the claim of Step 4, we consider any function $p$ such that (48) holds. We define a new price function $\overline{p}$ as follows:

$$\overline{p}(f) = \begin{cases} fp(1), & 0 \leq f \leq 1; \\ p(1) + \dfrac{\partial^+ p(1)}{\partial f}(f-1), & f \geq 1. \end{cases} \quad (53)$$

(See Figure 1 for an illustration.) Let $a = p(1)$, and let $b = \partial^+ p(1)/\partial f$. Then $a > 0$; and since $p(0) = 0$, we have $\partial^+ p(1)/\partial f \geq p(1)$ by convexity of $p$, so that $b \geq a$. Furthermore, since $p(1) < 1$ from (48), we have $1/a > 1$. Finally, we have:

$$\frac{1}{a+b} = \frac{1}{p(1)}(1 - \beta^+(1)) \leq 1,$$

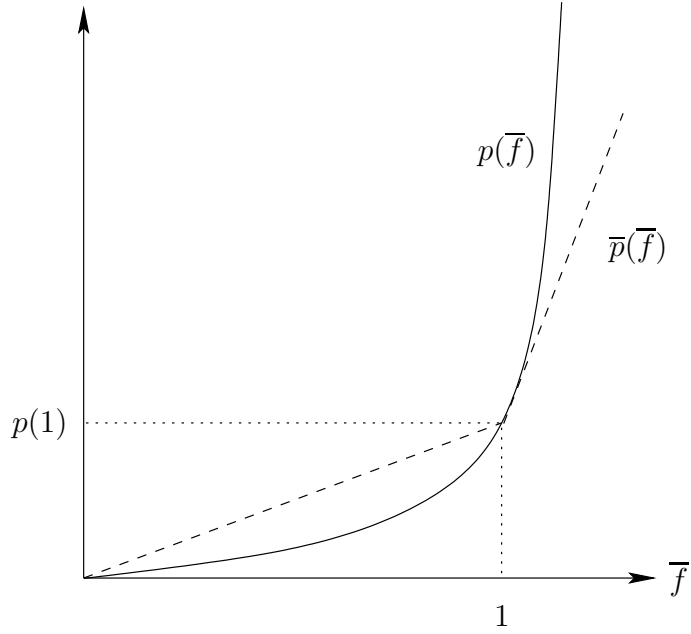

Figure 1: Proof of Theorem 8, Step 4: Given a price function $p$ (solid line) and Nash equilibrium rate 1, a new price function $\overline{p}$ (dashed line) is defined according to (53).

where the equality follows from the definition of $\beta^+(1)$ and the inequality follows from (48). Thus $\overline{p}$ satisfies (50)-(52). Observe also that $\overline{p}(1) = p(1)$, and $\partial^+\overline{p}(1)/\partial f = \partial^+ p(1)/\partial f$, and thus $\overline{\beta}^+(1) = \beta^+(1)$.

We now show that $F(\overline{p}) \leq F(p)$. As an intermediate step, we define a new price function $\hat{p}(\cdot)$ as follows:
$$\hat{p}(f) = \begin{cases} p(f), & 0 \leq f \leq 1; \\ \overline{p}(f), & f \geq 1. \end{cases}$$
Of course, $\hat{p}(1) = p(1)$ and $\partial^+\hat{p}(1)/\partial f = \partial^+\overline{p}(1)/\partial f = \partial^+ p(1)/\partial f$, so that (48) is satisfied for $\hat{p}$. Let $\hat{C}(f) = \int_0^f \hat{p}(z)\,dz$ denote the cost function associated with $\hat{p}(\cdot)$. Observe that (by convexity of $p$), we have for all $f$ that $\hat{p}(f) \leq p(f)$, so that $\hat{C}(f) \leq C(f)$. Thus:
$$\max_{f \geq 0}(f - \hat{C}(f)) \geq \max_{f \geq 0}(f - C(f)).$$

Furthermore, $\hat{C}(1) = C(1)$ so that $F(\hat{p}) \leq F(p)$.

Next, we let $\overline{C}(f) = \int_0^f \overline{p}(z)\,dz$ denote the cost function associated with $\overline{p}(\cdot)$. By convexity of $p$, we know $\overline{p}(1) \geq p(1)$ for $0 \leq f \leq 1$; thus $\overline{C}(f) \geq C(f)$ in that region. We let $\Delta \triangleq$

$\overline{C}(1) - C(1) \geq 0$. Then we have the following relationship:

$$F(\hat{p}) = \frac{p(1) + (1-p(1))^2/\beta^+(1) - C(1)}{\max_{f \geq 0}\left(f - \hat{C}(f)\right)} \tag{54}$$

$$\geq \frac{p(1) + (1-p(1))^2/\beta^+(1) - (C(1) + \Delta)}{\max_{f \geq 0}\left(f - (\hat{C}(f) + \Delta)\right)} \tag{55}$$

$$= F(\overline{p}). \tag{56}$$

The last equality follows by observing that since $\hat{p}(1) = p(1) < 1$, the solution to $\max_{f \geq 0}(f - \hat{C}(f))$ occurs at $\hat{f}^S > 1$ where $\hat{p}(\hat{f}^S) = 1$; and at all points $f \geq 1$, we have the relationship $\hat{C}(f) + \Delta = \overline{C}(f)$. Combining the preceding results, we have $F(p) \geq F(\overline{p})$, as required.

*Step 5: The minimum value of $F(p)$ over all $p$ satisfying (50)-(52) is $4\sqrt{2} - 5$.* We first show that given $p$ satisfying (50)-(52), $F(p)$ is given by:

$$F(p) = \frac{\frac{1}{2}a + \left(1 + \frac{a}{b}\right)(1-a)^2}{1 - \frac{1}{2}a + \frac{1}{2}\frac{(1-a)^2}{b}} = \frac{ab + 2(a+b)(1-a)^2}{2b - ab + (1-a)^2}. \tag{57}$$

The numerator results by simplifying the numerator of (47), when $p$ takes the form described by (50)-(52). To arrive at the denominator, we note that the solution to $\max_{f \geq 0}(f - C(f))$ occurs at $f^S$ satisfying $p(f^S) = 1$. Since $a < 1$, we must have $f^S > 1$ and:

$$a + b(f^S - 1) = 1.$$

Simplifying, we find:

$$f^S = 1 + \frac{1-a}{b}. \tag{58}$$

The expression $f^S - C(f^S)$, upon simplification, becomes the denominator of (57), as required.

Fix $a$ and $b$ such that $0 < a \leq b$, and $1/(a+b) \leq 1 < 1/a$, and define $p$ as in (50). We note here that the constraints $0 < a \leq b$ and $1/(a+b) \leq 1 < 1/a$ may be equivalently rewritten as $0 < a < 1$, and $\max\{a, 1-a\} \leq b$. Define $H(a,b) \triangleq F(p)$; from (57), note that for fixed $a$, $H(a,b)$ is a ratio of two affine functions of $b$, and thus the minimal value of $H(a,b)$ is achieved either when $b = \max\{a, 1-a\}$ or as $b \to \infty$. Define $H_1(a) = H(a,b)|_{\max\{a,1-a\}}$, and $H_2(a) = \lim_{b \to \infty} H(a,b)$. Then:

$$H_1(a) = \begin{cases} H(a,b)|_{b=1-a} = \dfrac{2-a}{3-2a}, & \text{if } 0 < a \leq 1/2; \\[1em] H(a,b)|_{b=a} = a^2 + 4a(1-a)^2, & \text{if } 1/2 \leq a < 1; \end{cases} \tag{59}$$

$$H_2(a) = \lim_{b \to \infty} H(a,b) = \frac{a + 2(1-a)^2}{2-a}. \tag{60}$$

We now minimize $H_1(a)$ and $H_2(a)$ over $0 < a < 1$. Over $0 < a \leq 1/2$, the minimum value of $H_1(a)$ is $2/3$, achieved as $a \to 0$. Over $1/2 \leq a < 1$, the minimum value of $H_1(a)$ is $20/27$, achieved at $a = 2/3$. Finally, over $0 < a < 1$, the minimum value of $H_2(a)$ is $4\sqrt{2} - 5$, achieved at $a = 2 - \sqrt{2}$. Since $\min\{2/3, 20/27, 4\sqrt{2} - 5\} = 4\sqrt{2} - 5$, we conclude that the minimal value of $F(p)$ over all $p$ satisfying (50)-(52) is equal to $4\sqrt{2} - 5$. This completes the proof of (26), the lower bound in the theorem.

We now show that this lower bound is tight. Fix $\delta > 0$. The preceding argument shows that the worst case occurs for price functions satisfying (50)-(52), where $a = 2 - \sqrt{2}$ and $b \to \infty$. For fixed $b \geq a = 2 - \sqrt{2}$, let $p_b$ be the associated price function defined according to (50). Then we have established that:

$$\lim_{b \to \infty} F(p_b) = 4\sqrt{2} - 5.$$

From Lemma 9, we know there exists $\gamma_b$ such that $\gamma_b < F(p_b) + \delta/2$, and where $\gamma_b$ is the ratio of Nash equilibrium aggregate surplus to maximal aggregate surplus for some game with price function $p_b$ and total allocated rate 1 at the Nash equilibrium. We thus have:

$$\lim_{b \to \infty} \gamma_b = \lim_{b \to \infty} F(p_b) + \delta/2 = 4\sqrt{2} - 5 + \delta/2.$$

Thus for $b$ sufficiently large, we will have $\gamma_b < 4\sqrt{2} - 5 + \delta$, establishing (27). $\square$

Theorem 8 shows that in the worst case, aggregate surplus falls by no more than approximately 34% when users are able to anticipate the effects of their actions on the price of the link. Furthermore, this bound is essentially tight. In fact, from the proof of the theorem we see that this ratio is achieved via a sequence of games where:

1. The price function $p$ has the form given by (50)-(52), with $a = 2 - \sqrt{2}$, $b \to \infty$, and $f = 1$;

2. The number of users becomes large ($R \to \infty$); and

3. User 1 has linear utility with $U_1(d_1) = d_1$, and all users have linear utility with $U_r(d_r) = \alpha_r d_r$, where $\alpha_r \approx p(1) = 2 - \sqrt{2}$ (for $r > 1$).

The last item follows by substituting the solution (45) in (37), and taking the limit as $R \to \infty$. (Note that formally, we must take care that the limits of $R \to \infty$ and $b \to \infty$ are taken in the correct order; in particular, in the proof we first have $R \to \infty$, and then $b \to \infty$.)

Note that the price function $p$ used to achieve the worst case efficiency loss is not differentiable. As discussed in Section 1, this is the main reason that we allow nondifferentiable price functions in Assumption 2. Indeed, some of the results of Section 2.1 can be simplified if we restrict attention only to differentiable price functions.

It is interesting to note that the worst case is obtained by considering instances where the price function is becoming steeper and steeper at the Nash equilibrium rate 1, since $b \to \infty$. This forces the optimal rate $f^S$ at the solution to *SYSTEM* to approach the Nash equilibrium rate $f = 1$, as we observe from (58); nevertheless, the shortfall between the Nash equilibrium aggregate surplus and the maximal aggregate surplus approaches 34%.

# 4 Inelastic Supply vs. Elastic Supply

In this section we briefly compare the model of this paper (allocation of a resource in elastic supply) with the model of [12] (allocation of a resource in inelastic supply). In [12], a model is considered with a single link having exactly $C$ units of rate available to allocate among the users. As in the model of this paper, user $r$ submits a bid $w_r$. The link manager then sets a price $\mu = \sum_r w_r / C$; and user $r$ receives an allocation $d_r$ given by:

$$d_r = \begin{cases} 0, & \text{if } w_r = 0; \\ \dfrac{w_r}{\mu}, & \text{if } w_r > 0. \end{cases}$$

As in this paper, the payoff to user $r$ is $U_r(d_r) - w_r$. It is shown in [12] that when users are price anticipating and the link supply is inelastic, the efficiency loss is at most 25% of the maximal aggregate utility.

Intuitively, we would like to model a system with an inelastic supply $C$ by a cost function which is zero for $0 \leq f < C$, and infinite for $f > C$. Formally, we show in this section that if the price function is given by $p(f) = af^B$ for $a \geq 0$ and $B \geq 1$, then as $B \to \infty$ the worst case efficiency loss approaches 25%—the same value obtained in [12]. While this does not formally establish the result in [12], the limit is intuitively plausible, because as the exponent $B$ increases, the price function $p$ and associated cost function begin to resemble an inelastic capacity constraint with $C = 1$: for $f < 1$, $f^B \to 0$ as $B \to \infty$; and for $f > 1$, $f^B \to \infty$ as $B \to \infty$.

**Theorem 10** *Suppose that Assumptions 1-3 hold. Suppose also that $U_r(0) \geq 0$ for all $r$, and that $p(f) = af^B$ for $a \geq 0$ and $B \geq 1$. Define the function $g(B)$ by:*

$$g(B) \triangleq \left( \frac{B+1}{2B+1} \right)^{1/B} \left( \frac{(B+1)(3B+2)}{(2B+1)^2} \right). \tag{61}$$

*If $\mathbf{d}^S$ is any solution to SYSTEM, and $\mathbf{w}$ is any Nash equilibrium of the game defined by $(Q_1, \ldots, Q_R)$, then:*

$$\text{surplus}(\mathbf{d}(\mathbf{w})) \geq g(B) \cdot \text{surplus}(\mathbf{d}^S), \tag{62}$$

*where $\text{surplus}(\cdot)$ is defined in (3). Furthermore, $g(B)$ is strictly increasing, with $g(B) \to 3/4$ as $B \to \infty$; and the bound (62) is tight: for fixed $B \geq 1$, for every $\delta > 0$, there exists a choice of $R$ and a choice of (linear) utility functions $U_r$, $r = 1, \ldots, R$, such that a Nash equilibrium $\mathbf{w}$ a solution $\mathbf{d}^S$ to SYSTEM exist with:*

$$\text{surplus}(\mathbf{d}(\mathbf{w})) \leq (g(B) + \delta) \cdot \text{surplus}(\mathbf{d}^S). \tag{63}$$

*Proof.* We follow the proof of Theorem 8. Steps 1-4 follow as in that proof, provided we can show that two scalings of the function $p(\cdot)$ do not affect our result—in Step 1, where we replace $p(\cdot)$ by $p(\cdot)/\max_r \alpha_r$, and in Step 2, where we replace $p(\cdot)$ by $p(f \cdot)$, where $f$ is the Nash equilibrium rate. Indeed, both these scalings remain valid, since the rescaled price function is still a monomial with the same exponent as $p$, but a different constant coefficient. In particular, we may continue to restrict attention to the special case where $U_r(d_r) = \alpha_r d_r$, with $\max_r \alpha_r = \alpha_1 = 1$, and where the total Nash equilibrium allocated rate is 1.

From Steps 1-4 of the proof of Theorem 8, we must minimize $F(p)$, defined in (47), for all choices of $p$ such that (48) is satisfied, i.e., such that:

$$1 - \beta^+(1) \leq p(1) < 1.$$

For $p(f) = af^B$, we have $\beta^+(f) = B/(1+B)$; and thus we require:

$$\frac{1}{1+B} \leq a < 1. \tag{64}$$

Note that at the maximal aggregate surplus, $p(f^S) = a(f^S)^B = 1$ implies that $f^S = a^{-1/B}$. Furthermore, $C(f) = af^{B+1}/(B+1)$ for $f \geq 0$. Thus $f^S - C(f^S)$ is given by:

$$f^S - C(f^S) = \left(\frac{B}{B+1}\right) \cdot \left(\frac{1}{a}\right)^{1/B}.$$

From (47), we conclude that $F(p)$ is given by:

$$F(p) = \frac{a + (1-a)^2(1+1/B) - a/(B+1)}{\left(\frac{B}{B+1}\right)\left(\frac{1}{a}\right)^{1/B}}.$$

We now minimize $F(p)$ over the set of $a$ satisfying (64). We begin by differentiating $F(p)$ with respect to $a$, and setting the derivative to zero; simplifying, this yields the following equation:

$$Ba + \left(1 + \frac{1}{B}\right)\left((2B+1)a^2 - 2(B+1)a + 1\right) = 0.$$

This equation is quadratic in $a$, and has two solutions $a_1$ and $a_2$: $a_1 = 1/(B+1)$, and $a_2 = (B+1)/(2B+1)$. Both solutions satisfy (64). Let $p_1(f) = a_1 f^B$, and $p_2(f) = a_2 f^B$. We have:

$$F(p_1) = \left(\frac{1}{B+1}\right)^{1/B}\left(\frac{B+2}{B+1}\right); \quad F(p_2) = \left(\frac{B+1}{2B+1}\right)^{1/B}\left(\frac{(B+1)(3B+2)}{(2B+1)^2}\right).$$

To minimize $F(p)$ over $a$ satisfying (64), we need also to check the endpoint where $a = 1$. If $p = f^B$, we find $F(p) = 1$; since $F(p_1), F(p_2) \leq 1$ from the definition of $F(p)$, the minimum value is achieved at either $p_1$ or $p_2$.

For $B \geq 1$, we define $g_1(B) = F(p_1)$, and $g_2(B) = F(p_2)$. We need the following technical lemma.

**Lemma 11** *The functions $g_1(B)$ and $g_2(B)$ are strictly increasing for $B \geq 1$. Furthermore, $g_1(B) \geq 3/4$ for $B \geq 1$, while $\lim_{B \to \infty} g_2(B) = 3/4$.*

*Proof.* We begin by noting that $g_1(1) = 3/4$. Let $\hat{g}_1(B) = \ln(g_1(B))$; it suffices to show that $\hat{g}_1(B)$ is strictly increasing for $B \geq 1$. Differentiating $\hat{g}_1$ yields:

$$\hat{g}_1'(B) = \frac{(B+2)\ln(B+1) - 2B}{(B+2)B^2}.$$

It suffices to check that $h_1(B) > 0$, where:

$$h_1(B) = (B+2)\ln(B+1) - 2B.$$

We have $h_1(1) = 3\ln 2 - 2 > 0$; $h_1'(1) = \ln 2 - 1/2 > 0$; and $h_1''(B) = B/(B+1)^2 > 0$. This implies $h_1(B) > 0$ for all $B \geq 1$, so $g_1$ is strictly increasing for $B \geq 1$.

Next we consider $g_2(B)$. Note first that $g_2(1) = 20/27$. Furthermore, as $B \to \infty$, $((B+1)/(2B+1))^{1/B} \to 1$, and $(B+1)(3B+2)/(2B+1)^2 \to 3/4$. Thus $g_2(B) \to 3/4$ as $B \to \infty$.

Finally, let $\hat{g}_2(B) = \ln(g_2(B))$; it suffices to show $\hat{g}_2$ is strictly increasing for $B \geq 1$. Differentiating $\hat{g}_2(B)$ yields:

$$\hat{g}_2'(B) = \frac{(3B+2)\ln\left(\frac{2B+1}{B+1}\right) - 2B}{(3B+2)B^2}.$$

As above, it suffices to check that $h_2(B) > 0$, where:

$$h_2(B) = (3B+2)\ln\left(\frac{2B+1}{B+1}\right) - 2B.$$

We have $h_2(1) = 5\ln(3/2) - 2 > 0$; $h_2'(1) = 3\ln(3/2) - 7/6 > 0$; and $h_2''(1) = B/[(B+1)^2(2B+1)^2] > 0$. Thus $h_2(B) > 0$ for all $B \geq 1$, which implies $g_2$ is strictly increasing for $B \geq 1$. □

From the previous lemma, we conclude that the minimum value of $F(p)$ over $p = af^B$ satisfying (64) is given by $g_2(B)$; this establishes (62). As in Theorem 8, by construction this bound is tight, so (63) holds as well. □

The preceding theorem shows that for a particular sequence of price functions which approach an inelastic supply constraint, the efficiency loss gradually decreases from $7/27$ (at $B = 1$) to $1/4$ (as $B \to \infty$). In the limit as $B \to \infty$, we recover the same efficiency loss as in the earlier work of [12]. However, while we have demonstrated such a limit holds as long as the price functions are monomials, there remains an open question: if the price functions "converge" (in an appropriate sense) to a fixed capacity constraint, under what conditions does the efficiency loss also converge to 1/4? It is straightforward to check that such a limit cannot always hold. For example, consider price functions $p$ of the form specified in (50)-(52). Using the expression for $F(p)$ given in (57), it is possible to show that by first taking $b \to \infty$, and then taking $a \to 0$, the worst case efficiency loss approaches zero; see (60).

## 5 General Networks

In this section we consider an extension of the single link model to general networks, using methods similar to the network model presented in [12]. We consider a network consisting of $J$ links, or *resources*, numbered $1, \ldots, J$. As before, a set of users numbered $1, \ldots, R$, shares this network of resources. We assume there exists a set of paths through the network, numbered $1, \ldots, P$. By an abuse of notation, we use $J$, $R$, and $P$ to also denote the sets of resources, users, and paths, respectively. Each path $q \in P$ uses a subset of the set of resources $J$; if resource $j$ is used by path $q$, we denote this by writing $j \in q$. Each user $r \in R$ has a collection of paths available through the network; if path $q$ serves user $r$, we denote this by writing $q \in r$. We assume without loss of generality that paths are uniquely identified with users, so that for each path $q$ there exists a unique user $r$ such that $q \in r$. (There is no loss of generality because if two users share the same path, that is captured in our model by creating two paths which use exactly the same subset of resources.) For notational convenience, we note that the resources required by individual paths are captured by the *path-resource incidence matrix* $\mathbf{A}$, defined by:

$$A_{jq} = \begin{cases} 1, & \text{if } j \in q \\ 0, & \text{if } j \notin q. \end{cases}$$

Furthermore, we can capture the relationship between paths and users by the *path-user incidence matrix* $\mathbf{H}$, defined by:

$$H_{rq} = \begin{cases} 1, & \text{if } q \in r \\ 0, & \text{if } q \notin r. \end{cases}$$

Note that by our assumption on paths, for each path $q$ we have $H_{rq} = 1$ for exactly one user $r$.

Let $y_q \geq 0$ denote the rate allocated to path $q$, and let $d_r = \sum_{q \in r} y_q \geq 0$ denote the rate allocated to user $r$; using the matrix $\mathbf{H}$, we may write the relation between $\mathbf{d} = (d_r, r \in R)$ and $\mathbf{y} = (y_q, q \in P)$ as $\mathbf{Hy} = \mathbf{d}$. Furthermore, if we let $f_j$ denote the total rate on link $j$, we must have:

$$\sum_{q:j \in q} y_q = f_j, \quad j \in J.$$

Using the matrix $\mathbf{A}$, we may write this constraint as $\mathbf{Ay} = \mathbf{f}$.

We continue to assume that user $r$ receives a utility $U_r(d_r)$ from an allocated rate $d_r$, and that each link $j$ incurs a cost $C_j(f_j)$ when the total allocated rate at link $j$ is $f_j$. We make the following assumptions regarding the utility functions and cost functions.

**Assumption 5** *For each $r$, the utility function $U_r(d_r)$ is concave, nondecreasing, and continuous over the domain $d_r \geq 0$.*

**Assumption 6** *For each $j$, there exists a continuous, convex, strictly increasing function $p_j(f_j)$ over $f_j \geq 0$ with $p_j(0) = 0$, such that for $f_j \geq 0$:*

$$C_j(f_j) = \int_0^{f_j} p_j(z)dz.$$

*Thus $C_j(f_j)$ is strictly convex and increasing.*

Assumption 5 is similar to Assumption 1, but we no longer require that $U_r$ be strictly increasing or differentiable. For this reason the results of this section are also generalizations of the corresponding results of Sections 1 and 3 for a single link. Assumption 6 is identical to Assumption 2, for each link $j$.

The natural generalization of the problem *SYSTEM* to a network context is given by the following optimization problem:

*SYSTEM*:

$$\text{maximize} \quad \sum_r U_r(d_r) - \sum_j C_j(f_j) \tag{65}$$

$$\text{subject to} \quad \mathbf{Ay} = \mathbf{f}; \tag{66}$$

$$\mathbf{Hy} = \mathbf{d}; \tag{67}$$

$$y_q \geq 0, \quad q \in P. \tag{68}$$

Since the objective function is continuous and $U_r$ grows at most linearly while $C_j$ grows superlinearly, an optimal solution $\mathbf{y}$ exists. Since the feasible region is convex and the cost functions $C_j$ are each strictly convex, the optimal vector $\mathbf{f} = \mathbf{Ay}$ is uniquely defined (though $\mathbf{y}$ need not be unique). In addition, if the functions $U_r$ are strictly concave, then the optimal vector $\mathbf{d} = \mathbf{Hy}$ is uniquely defined as well. We continue to refer to the objective function (65) as the *aggregate surplus*, and modify our definition of surplus$(\cdot)$ accordingly, in terms of the allocation $\mathbf{d}$ made to the users and the aggregate rate $\mathbf{f}$ allocated by the links:

$$\text{surplus}(\mathbf{d}, \mathbf{f}) \triangleq \sum_r U_r(d_r) - \sum_j C_j(f_j). \tag{69}$$

As in the previous development, we use the solution to *SYSTEM* as a benchmark for the outcome of the network game.

We now define the resource allocation mechanism for this network setting. The natural extension of the single link model is defined as follows. Each user $r$ submits a *bid* $w_{jr}$ for each resource $j$; this defines a strategy for user $r$ given by $\mathbf{w}_r = (w_{jr}, j \in J)$, and a composite strategy vector given by $\mathbf{w} = (\mathbf{w}_1, \ldots, \mathbf{w}_R)$. We then assume that each link takes these bids as input, and uses the pricing scheme developed in the Section 1. This is formalized in the following assumption, which is a direct analogue of Assumption 3 for each link $j$.

**Assumption 7** *For all $\mathbf{w} \geq 0$, at each link $j$ the aggregate rate $f_j(\mathbf{w})$ is the solution $f_j$ to:*

$$\sum_r w_{jr} = f_j p_j(f_j). \tag{70}$$

*Furthermore, for each $r$, $x_{jr}(\mathbf{w})$ is given by:*

$$x_{jr}(\mathbf{w}) = \begin{cases} 0, & \text{if } w_{jr} = 0; \\ \dfrac{w_{jr}}{p_j(f_j(\mathbf{w}))}, & \text{if } w_{jr} > 0. \end{cases} \tag{71}$$

We define the vector $\mathbf{x}_r(\mathbf{w})$ by:

$$\mathbf{x}_r(\mathbf{w}) = (x_{jr}(\mathbf{w}), j \in J).$$

Now given any vector $\overline{\mathbf{x}}_r = (\overline{x}_{jr}, j \in J)$, we define $d_r(\overline{\mathbf{x}}_r)$ to be the optimal objective value of the following optimization problem:

$$\text{maximize} \quad \sum_{q \in r} y_q \tag{72}$$

$$\text{subject to} \quad \sum_{q \in r : j \in p} y_q \leq \overline{x}_{jr}, \quad j \in J; \tag{73}$$

$$y_q \geq 0, \quad q \in r. \tag{74}$$

Given the strategy vector $\mathbf{w}$, we define the rate allocated to user $r$ as $d_r(\mathbf{x}_r(\mathbf{w}))$. To gain some intuition for this allocation mechanism, notice that when the vector of bids is $\mathbf{w}$, user $r$ is allocated a rate $x_{jr}(\mathbf{w})$ at each link $j$. Since the utility to user $r$ is nondecreasing in the total amount of rate allocated, user $r$'s utility is maximized if he solves the preceding optimization problem with $\overline{\mathbf{x}}_r = \mathbf{x}_r(\mathbf{w})$, which is a *max-flow* problem constrained by the rate $x_{jr}(\mathbf{w})$ available at each link $j$. In other words, user $r$ is allocated the maximum possible rate $d_r(\mathbf{x}_r(\mathbf{w}))$, given that each link $j$ has granted him rate $x_{jr}(\mathbf{w})$.

As before we adopt the notation $\mathbf{w}_{-r} = (\mathbf{w}_1, \ldots, \mathbf{w}_{r-1}, \mathbf{w}_{r+1}, \ldots, \mathbf{w}_R)$. Based on the definition of $d_r(\mathbf{x}_r(\mathbf{w}))$ above, the payoff to user $r$ is given by:

$$Q_r(\mathbf{w}_r; \mathbf{w}_{-r}) = U_r\big(d_r(\mathbf{x}_r(\mathbf{w}))\big) - \sum_j w_{jr}. \tag{75}$$

A *Nash equilibrium* of the game defined by $(Q_1, \ldots, Q_R)$ is a vector $\mathbf{w} \geq 0$ such that for all $r$:

$$Q_r(\mathbf{w}_r; \mathbf{w}_{-r}) \geq Q_r(\overline{\mathbf{w}}_r; \mathbf{w}_{-r}), \quad \text{for all } \overline{\mathbf{w}}_r \geq 0. \tag{76}$$

As in the development of Section 2.1, the following proposition plays a key role in demonstrating existence of a Nash equilibrium. The proof is identical to the proof of Proposition 3, and is omitted.

**Proposition 12** *Suppose that Assumptions 5-7 hold. Then for each $j$ and each $r$: (1) $x_{jr}(\mathbf{w})$ is a continuous function of $\mathbf{w}$; and (2) for any $\mathbf{w}_{-r} \geq 0$, $x_{jr}(\mathbf{w})$ is strictly increasing and concave in $w_{jr} \geq 0$, and $x_{jr}(\mathbf{w}) \to \infty$ as $w_{jr} \to \infty$.*

As in Proposition 4, the following proposition gives existence of a Nash equilibrium for the game defined by $(Q_1, \ldots, Q_R)$.

**Proposition 13** *Suppose that Assumptions 5-7 hold. Then there exists a Nash equilibrium $\mathbf{w}$ for the game defined by $(Q_1, \ldots, Q_R)$.*

*Proof.* The proof follows the proof of Proposition 4. The only step which requires modification is to show that the payoff $Q_r$ of user $r$ is a concave function of $\mathbf{w}_r$ and a continuous function of $\mathbf{w}$. To prove this, it suffices to show that $U_r(d_r(\mathbf{x}_r(\mathbf{w}_r; \mathbf{w}_{-r})))$ is a concave function of $\mathbf{w}_r$ and a continuous function of $\mathbf{w}$. We first observe that by Proposition 12, $x_{jr}(\mathbf{w})$ is a concave function of $w_{jr} \geq 0$, and a continuous function of $\mathbf{w}$. Since for each $j$ the function $x_{jr}(\mathbf{w})$ does not depend on $w_{kr}$ for $k \neq j$, we conclude that each component of $\mathbf{x}_r(\mathbf{w}_r; \mathbf{w}_{-r})$ is a concave function of $\mathbf{w}_r$. Now since $d_r$ is defined as the optimal objective value of a linear program, $d_r(\overline{\mathbf{x}}_r)$ is continuous and concave as a function of $\overline{\mathbf{x}}_r$ [4]. In addition, $d_r(\overline{\mathbf{x}}_r)$ is *nondecreasing* in $\overline{\mathbf{x}}_r$; i.e., if $\overline{x}_{jr} \geq \hat{x}_{jr}$ for all $j$, then $d_r(\overline{\mathbf{x}}_r) \geq d_r(\hat{\mathbf{x}}_r)$ (this follows from the problem (72)-(74)). These properties of $x_{jr}$ and $d_r$, combined with the fact that $U_r$ is concave, continuous, and nondecreasing from Assumption 5, imply that $U_r(d_r(\mathbf{x}_r(\mathbf{w}_r; \mathbf{w}_{-r})))$ is a concave function of $\mathbf{w}_r$, and a continuous function of $\mathbf{w}$. □

The following theorem demonstrates that the utility lost at any Nash equilibrium is no worse than $4\sqrt{2}-5$ of the maximum possible aggregate surplus, matching the result derived for the single link model. The proof follows the proof of Theorem 9 in [12]: we construct a single link game at each link $j$, whose Nash equilibrium is the same as the fixed Nash equilibrium of the network game. We then apply Theorem 8 at each link to complete the proof. However, we note that this result does not require $U_r$ to be strictly increasing or continuously differentiable, and is therefore a stronger version of Theorem 8 for the single link case.

**Theorem 14** *Suppose that Assumptions 5-7 hold. Assume also that $U_r(0) \geq 0$ for all users $r$. Let $\mathbf{w}$ be any Nash equilibrium for the game defined by $(Q_1, \ldots, Q_R)$, and let $(\mathbf{y}^S, \mathbf{f}^S, \mathbf{d}^S)$ be any solution to SYSTEM. If we define $d_r^{NE} = d_r(\mathbf{x}_r(\mathbf{w}))$ for each $r$, and $f_j^{NE} = f_j(\mathbf{w})$ for each $j$, then:*
$$\mathrm{surplus}(\mathbf{d}^{NE}, \mathbf{f}^{NE}) \geq (4\sqrt{2} - 5) \cdot \mathrm{surplus}(\mathbf{d}^S, \mathbf{f}^S),$$
*where* $\mathrm{surplus}(\cdot, \cdot)$ *is defined in* (69).

*Proof.* The proof consists of three main steps. First, we describe the entire problem in terms of the vector $\mathbf{x}_r(\mathbf{w}) = (x_{jr}(\mathbf{w}), j \in J)$ of the rate allocations to user $r$ from the network. We show in Lemma 15 that Nash equilibria can be characterized in terms of each user $r$ optimally choosing a rate allocation $\overline{\mathbf{x}}_r = (\overline{x}_{jr}, j \in J)$, given the vector of bids $\mathbf{w}_{-r}$ of all other users.

In the second step, we observe that the utility to user $r$ given a vector of rate allocations $\overline{\mathbf{x}}_r$ is exactly $U_r(d_r(\overline{\mathbf{x}}_r))$; we call this a "composite" utility function. In Lemma 16, we linearize this composite utility function; formally, we replace $U_r(d_r(\overline{\mathbf{x}}_r))$ with a linear function $\boldsymbol{\alpha}_r^\top \overline{\mathbf{x}}_r$. The difficulty in this phase of the analysis is that the composite utility function $U_r(d_r(\cdot))$ may not be differentiable, because the max-flow function $d_r(\cdot)$ is not differentiable everywhere; as a result, convex analytic techniques are required.

Finally, we conclude the proof by observing that when the "composite" utility function for user $r$ is linear in the vector of rate allocations $\overline{\mathbf{x}}_r$, the network structure is no longer relevant. In this case the game defined by $(Q_1, \ldots, Q_R)$ decouples into $J$ games, one for each link. We then apply Theorem 8 at each link to arrive at the bound in the theorem.

We start by describing the entire problem in terms of the vector $\mathbf{x}_r(\mathbf{w}) = (x_{jr}(\mathbf{w}), j \in J)$ of the rate allocations to user $r$ from the network. We begin by redefining the problem *SYSTEM* as

follows:

$$\text{maximize} \quad \sum_r U_r(d_r(\overline{\mathbf{x}}_r)) - \sum_j C_j(f_j) \tag{77}$$

$$\text{subject to} \quad \sum_r \overline{x}_{jr} = f_j, \quad j \in J; \tag{78}$$

$$\overline{x}_{jr} \geq 0, \quad j \in J, r \in R. \tag{79}$$

(The notation $\overline{\mathbf{x}}_r$ is used here to distinguish from the function $\mathbf{x}_r(\mathbf{w})$.) In this problem, the network only chooses how to allocate rate at each link to the users. The users then solve a max-flow problem to determine the maximum rate at which they can send (this is captured by the function $d_r(\cdot)$). This problem is equivalent to the problem *SYSTEM* as defined in (65)-(68), because of the definition of $d_r(\cdot)$ in (72)-(74). We label an optimal solution to this problem by $(\mathbf{x}_r^S, r \in R; f_j^S, j \in J)$.

Our next step is to show that a Nash equilibrium may be characterized in terms of users optimally choosing rate allocations $(\overline{\mathbf{x}}_r, r \in R)$. We begin by "inverting" the function $x_{jr}(\mathbf{w})$, with respect to $w_{jr}$; that is, we determine the amount that user $r$ must pay to link $j$ to receive a predetermined rate allocation $\overline{x}_{jr}$, given that all other users have bid $\mathbf{w}_{-r}$. Formally, we observe from Proposition 12 that $x_{jr}(\mathbf{w})$ is concave, strictly increasing, and continuous in $w_{jr}$. Finally, since $x_{jr}(\mathbf{w}) = 0$ if $w_{jr} = 0$, and $x_{jr}(\mathbf{w}) \to \infty$ as $w_{jr} \to \infty$, we can define a function $\omega_{jr}(\overline{x}_{jr}; \mathbf{w}_{-r})$ for $\overline{x}_{jr} \geq 0$, which satisfies:

$$x_{jr}(\mathbf{w}) = \overline{x}_{jr} \quad \text{if and only if} \quad w_{jr} = \omega_{jr}(\overline{x}_{jr}; \mathbf{w}_{-r}).$$

From the properties of $x_{jr}$ described above, we note that for a fixed vector $\mathbf{w}_{-r}$, the function $\omega_{jr}(\cdot; \mathbf{w}_{-r})$ is convex, strictly increasing, and continuous, with $\omega_{jr}(0; \mathbf{w}_{-r}) = 0$ and $\omega_{jr}(\overline{x}_{jr}; \mathbf{w}_{-r}) \to \infty$ as $\overline{x}_{jr} \to \infty$.

We now use the functions $\omega_{jr}$ to write user $r$'s payoff in terms of the allocated rate vector $\overline{\mathbf{x}}_r = (\overline{x}_{jr}, j \in J)$, rather than in terms of the bid $\mathbf{w}_r$. For $\overline{\mathbf{x}}_r \geq 0$, we define a function $F_r(\overline{\mathbf{x}}_r; \mathbf{w}_{-r})$ as follows:

$$F_r(\overline{\mathbf{x}}_r; \mathbf{w}_{-r}) \triangleq U_r(d_r(\overline{\mathbf{x}}_r)) - \sum_j \omega_{jr}(\overline{x}_{jr}; \mathbf{w}_{-r}). \tag{80}$$

We now have the following lemma, which shows a Nash equilibrium may be characterized by an optimal choice of $\overline{\mathbf{x}}_r$ for each $r$.

**Lemma 15** *A vector $\mathbf{w}$ is a Nash equilibrium if and only if the following condition holds for each user $r$:*

$$\mathbf{x}_r(\mathbf{w}) \in \arg\max_{\overline{\mathbf{x}}_r \geq 0} F_r(\overline{\mathbf{x}}_r; \mathbf{w}_{-r}). \tag{81}$$

*Proof of Lemma.* Fix a bid vector $\mathbf{w}$, and suppose that there exists a vector $\overline{\mathbf{x}}_r \geq 0$ such that:

$$F_r(\overline{\mathbf{x}}_r; \mathbf{w}_{-r}) > F_r(\mathbf{x}_r(\mathbf{w}); \mathbf{w}_{-r}). \tag{82}$$

Since $\omega_{jr}(x_{jr}(\mathbf{w}); \mathbf{w}_{-r}) = w_{jr}$, we have $F_r(\mathbf{x}_r(\mathbf{w}); \mathbf{w}_{-r}) = Q_r(\mathbf{w}_r; \mathbf{w}_{-r})$. Now consider the bid vector $\overline{\mathbf{w}}_r$ defined by $\overline{w}_{jr} = \omega_{jr}(\overline{x}_{jr}; \mathbf{w}_{-r})$. Then $x_{jr}(\overline{\mathbf{w}}_r; \mathbf{w}_{-r}) = \overline{x}_{jr}$ for each $j$, so:

$$Q_r(\overline{\mathbf{w}}_r; \mathbf{w}_{-r}) = F_r(\overline{\mathbf{x}}_r; \mathbf{w}_{-r}).$$

Thus $\overline{\mathbf{w}}_r$ is a profitable deviation for user $r$, so $\mathbf{w}$ could not have been a Nash equilibrium.

Conversely, suppose that $\mathbf{w}$ is not a Nash equilibrium. As above, we have $F_r(\mathbf{x}_r(\mathbf{w}); \mathbf{w}_{-r}) = Q_r(\mathbf{w}_r; \mathbf{w}_{-r})$. Fix a user $r$, and let $\overline{\mathbf{w}}_r$ be a profitable deviation for user $r$, so that $Q_r(\overline{\mathbf{w}}_r; \mathbf{w}_{-r}) > Q_r(\mathbf{w}_r; \mathbf{w}_{-r})$. For each $j$, let $\overline{x}_{jr} = x_{jr}(\overline{\mathbf{w}}_r; \mathbf{w}_{-r})$. Then $\omega_{jr}(\overline{x}_{jr}; \mathbf{w}_{-r}) = w_{jr}$, so that $F_r(\overline{\mathbf{x}}_r; \mathbf{w}_{-r}) = Q_r(\overline{\mathbf{w}}_r; \mathbf{w}_{-r})$. Thus we have $F_r(\overline{\mathbf{x}}_r; \mathbf{w}_{-r}) > F_r(\mathbf{x}_r(\mathbf{w}); \mathbf{w}_{-r})$, so that (81) does not hold. □

Now suppose that $\mathbf{w}$ is a Nash equilibrium. Our approach is to replace user $r$ by $J$ users (which we call "virtual" users), one at each link $j$; this process has the effect of *isolating* each of the links, and removes any dependence on network structure. We define the virtual users so that $\mathbf{w}$ remains a Nash equilibrium at each single link game. Formally, for each user $r$, we construct a vector $\boldsymbol{\alpha}_r = (\alpha_{jr}, j \in J)$, and consider a single link game at each link $j$ where user $r$ has linear utility function $U_{jr}(x_{jr}) = \alpha_{jr} x_{jr}$. We choose the vectors $\boldsymbol{\alpha}_r$ so that the Nash equilibrium at each single link game is also given by $\mathbf{w}$; we then apply the result of Theorem 8 for the single link model to complete the proof of the theorem.

We now proceed to construct this vector $\boldsymbol{\alpha}_r$. Before continuing, we extend the notion of subgradients developed earlier to functions over $\mathbb{R}^J$. An *extended real-valued function* is a function $g : \mathbb{R}^J \to [-\infty, \infty]$; such a function is called *proper* if $g(x) > -\infty$ for all $x$, and $g(x) < \infty$ for at least one $x$. We say that a vector $\boldsymbol{\gamma} \in \mathbb{R}^J$ is a *subgradient* of an extended real-valued function $g$ at $\mathbf{x}$ if for all $\overline{\mathbf{x}} \in \mathbb{R}^J$, we have $g(\overline{\mathbf{x}}) \geq g(\mathbf{x}) + \boldsymbol{\gamma}^\top (\overline{\mathbf{x}} - \mathbf{x})$. The *subdifferential* of $g$ at $\mathbf{x}$, denoted $\partial g(\mathbf{x})$, is the set of all subgradients of $g$ at $\mathbf{x}$. For details on these concepts, we refer the reader to [23].

Lemma 15 allows us to characterize the Nash equilibrium $\mathbf{w}$ as a choice of optimal rate allocation $\overline{\mathbf{x}}_r$ by each user $r$, given the strategy vector $\mathbf{w}_{-r}$ of all other users. We recall the definition of $F_r$ in (80); we will now view $F_r$ as an extended real valued function, by defining $F_r(\overline{\mathbf{x}}_r; \mathbf{w}_{-r}) = -\infty$ if $\overline{x}_{jr} < 0$ for some $j$. We also define extended real-valued functions $G_r$ and $K_{jr}$ on $\mathbb{R}^J$ as follows:

$$G_r(\overline{\mathbf{x}}_r) = \begin{cases} U_r(d_r(\overline{\mathbf{x}}_r)), & \text{if } \overline{\mathbf{x}}_r \geq 0; \\ -\infty, & \text{otherwise.} \end{cases}$$

and

$$K_{jr}(\overline{\mathbf{x}}_r; \mathbf{w}_{-r}) = \begin{cases} -\omega_{jr}(\overline{x}_{jr}; \mathbf{w}_{-r}), & \text{if } \overline{x}_{jr} \geq 0; \\ 0, & \text{otherwise.} \end{cases}$$

Then we have $F_r = G_r + \sum_j K_{jr}$ on $\mathbb{R}^J$ (where we extend the definition of $+$ to $[-\infty, 0)$ in the obvious way). The following lemma establishes existence of the desired vector $\boldsymbol{\alpha}_r$.

**Lemma 16** *Let $\mathbf{w}$ be a Nash equilibrium. Then for each user $r$, there exists a vector $\boldsymbol{\alpha}_r = (\alpha_{jr}, j \in J) \geq 0$ such that $-\boldsymbol{\alpha}_r \in \partial[-G_r(\mathbf{x}_r(\mathbf{w}))]$, and the following relation holds:*

$$\mathbf{x}_r(\mathbf{w}) \in \arg\max_{\overline{\mathbf{x}}_r \geq 0} \left[ \boldsymbol{\alpha}_r^\top \overline{\mathbf{x}}_r - \sum_j \omega_{jr}(\overline{x}_{jr}; \mathbf{w}_{-r}) \right]. \tag{83}$$

*Proof of Lemma.* Fix a user $r$. We observe that $G_r$ is a concave function of $\overline{\mathbf{x}}_r \in \mathbb{R}^J$. This follows as in the proof of Proposition 13, because $d_r$ is a concave function of its argument (as it is the optimal objective value of the linear program (72)-(74)), and $U_r$ is nondecreasing and concave. Furthermore, we note that $K_{jr}(\overline{\mathbf{x}}_r; \mathbf{w}_{-r})$ is a concave function of $\overline{\mathbf{x}}_r \in \mathbb{R}^J$ as well, since $\omega_{jr}(\overline{x}_{jr}, \mathbf{w}_{-r})$ is convex and nonnegative for $\overline{x}_{jr} \geq 0$. Consequently, $F_r$ is a concave function of $\overline{\mathbf{x}}_r \in \mathbb{R}^J$. In particular, $-F_r$, $-G_r$, and $-K_{jr}$ are convex, proper extended real-valued functions. It is straightforward to show, using Theorem 23.8 in [23], that at $\mathbf{x}_r(\mathbf{w})$ we have:

$$\partial[-F_r(\mathbf{x}_r(\mathbf{w}); \mathbf{w}_{-r})] = \partial[-G_r(\mathbf{x}_r(\mathbf{w}))] + \sum_j \partial[-K_{jr}(\mathbf{x}_r(\mathbf{w}); \mathbf{w}_{-r})]. \tag{84}$$

(The summation here of the subdifferentials on the right hand side is a summation of sets, where $A + B = \{\mathbf{x} + \mathbf{y} : \mathbf{x} \in A, \mathbf{y} \in B\}$; if either $A$ or $B$ is empty, then $A + B$ is empty as well.)

Since $\mathbf{w}$ is a Nash equilibrium, from Lemma 15, we have for all $\overline{\mathbf{x}}_r \geq 0$ that:

$$F_r(\mathbf{x}_r(\mathbf{w}); \mathbf{w}_{-r}) \geq F_r(\overline{\mathbf{x}}_r; \mathbf{w}_{-r}).$$

Since $F_r(\overline{\mathbf{x}}_r; \mathbf{w}_{-r}) = -\infty$ if there exists $j$ such that $\overline{x}_{jr} < 0$, we in fact have $F_r(\mathbf{x}_r(\mathbf{w}); \mathbf{w}_{-r}) \geq F_r(\overline{\mathbf{x}}_r; \mathbf{w}_{-r})$ for *all* $\overline{\mathbf{x}}_r \in \mathbb{R}^J$ so we conclude $\mathbf{0}$ is a subgradient of $-F_r(\cdot; \mathbf{w}_{-r})$ at $\mathbf{x}_r(\mathbf{w})$. As a result, it follows from (84) that there exist vectors $\boldsymbol{\alpha}_r$ and $\boldsymbol{\beta}_{jr}$ with $-\boldsymbol{\alpha}_r \in \partial[-G_r(\mathbf{x}_r(\mathbf{w}))]$ and $-\boldsymbol{\beta}_{jr} \in \partial[-K_{jr}(\mathbf{x}_r(\mathbf{w}); \mathbf{w}_{-r})]$, such that $\boldsymbol{\alpha}_r = -\sum_j \boldsymbol{\beta}_{jr}$.

We first note that $G_r(\overline{\mathbf{x}}_r)$ is a nondecreasing function of $\overline{\mathbf{x}}_r$; that is, if $\overline{\mathbf{x}}_r \geq \hat{\mathbf{x}}_r$, then $G_r(\overline{\mathbf{x}}_r) \geq G_r(\hat{\mathbf{x}}_r)$. From this fact it follows that $\boldsymbol{\alpha}_r$ must be nonnegative, i.e., $\alpha_{jr} \geq 0$ for all $j$. It remains to show that (83) holds. We observe that $\mathbf{0} \in -\partial \hat{F}_r(\mathbf{x}_r(\mathbf{w}); \mathbf{w}_{-r})$, where $\hat{F}_r(\cdot; \mathbf{w}_{-r})$ is defined as follows:

$$\hat{F}_r(\overline{\mathbf{x}}_r; \mathbf{w}_{-r}) = \begin{cases} \boldsymbol{\alpha}_r^\top \overline{\mathbf{x}}_r - \sum_j \omega_{jr}(\overline{x}_{jr}; \mathbf{w}_{-r}), & \text{if } \overline{\mathbf{x}}_r \geq 0; \\ -\infty, & \text{otherwise.} \end{cases}$$

This observation follows by replacing $G_r(\overline{\mathbf{x}}_r)$ with the following function $\hat{G}_r$ on $\mathbb{R}^J$:

$$\hat{G}_r(\overline{\mathbf{x}}_r) = \begin{cases} \boldsymbol{\alpha}_r^\top \overline{\mathbf{x}}_r, & \text{if } \overline{\mathbf{x}}_r \geq 0; \\ -\infty, & \text{otherwise.} \end{cases}$$

Then we have $\hat{F}_r = \hat{G}_r + \sum_j K_{jr}$; and as before:

$$\partial[-\hat{F}_r(\mathbf{x}_r(\mathbf{w}); \mathbf{w}_{-r})] = \partial[-\hat{G}_r(\mathbf{x}_r(\mathbf{w}))] + \sum_j \partial[-K_{jr}(\mathbf{x}_r(\mathbf{w}); \mathbf{w}_{-r})].$$

The vector $-\boldsymbol{\alpha}_r$ is a subgradient of $-\hat{G}_r$ for all $\overline{\mathbf{x}}_r \geq 0$; in particular, $-\boldsymbol{\alpha}_r \in \partial[-\hat{G}_r(\mathbf{x}_r(\mathbf{w}))]$. We have already shown $\boldsymbol{\alpha}_r = -\sum_j \boldsymbol{\beta}_{jr} \in \sum_j \partial[-K_{jr}(\mathbf{x}_r(\mathbf{w}); \mathbf{w}_{-r})]$. Thus $\mathbf{0} \in \partial[-\hat{F}_r(\mathbf{x}_r(\mathbf{w}); \mathbf{w}_{-r})]$. This implies (83), as required. □

Let $\mathbf{w}$ be a Nash equilibrium. For each user $r$, fix the vector $\boldsymbol{\alpha}_r$ such that (83) holds. We start by observing that for each user $r$, since $-\boldsymbol{\alpha}_r$ is a subgradient of $-G_r(\mathbf{x}_r(\mathbf{w}))$, we have:

$$U_r(d_r(\mathbf{x}_r^S)) \leq U_r(d_r(\mathbf{x}_r(\mathbf{w}))) + \boldsymbol{\alpha}_r^\top (\mathbf{x}_r^S - \mathbf{x}_r(\mathbf{w})). \tag{85}$$

We distinguish two cases: either $\boldsymbol{\alpha}_r = \mathbf{0}$ for all $r$, or $\boldsymbol{\alpha}_r \neq \mathbf{0}$ for at least one $r$. We first consider the case where $\boldsymbol{\alpha}_r = \mathbf{0}$ for all $r$. Since $\omega_{jr}$ is strictly increasing, if $\boldsymbol{\alpha}_r = \mathbf{0}$, then the unique maximizer in (83) is $\overline{\mathbf{x}}_r = \mathbf{0}$. Thus, if $\boldsymbol{\alpha}_r = \mathbf{0}$ for all $r$, we must have $\mathbf{x}_r(\mathbf{w}) = \mathbf{0}$ for all $r$. But from (85), we have the following trivial inequality:

$$\sum_r U_r(d_r(\mathbf{x}_r(\mathbf{w}))) \geq \sum_r U_r(d_r(\mathbf{x}_r^S)).$$

Since $d_r(\mathbf{x}_r(\mathbf{w})) = d_r(\mathbf{0}) = 0$ for all $r$, this is only possible if $U_r(d_r(\mathbf{x}_r^S)) = U_r(0)$ for all $r$ as well. It follows that the aggregate surplus is zero at both the Nash equilibrium and the optimal solution to *SYSTEM*, so the theorem holds in this case.

We may assume without loss of generality, therefore, that $\boldsymbol{\alpha}_r \neq \mathbf{0}$ for at least one user $r$. We have the following simplification of (83):

$$\mathbf{x}_r(\mathbf{w}) \in \arg\max_{\overline{\mathbf{x}}_r \geq 0} \left[ \boldsymbol{\alpha}_r^\top \overline{\mathbf{x}}_r - \sum_j \omega_{jr}(\overline{x}_{jr}; \mathbf{w}_{-r}) \right]$$

$$= \arg\max_{\overline{\mathbf{x}}_r \geq 0} \left[ \sum_j (\alpha_{jr} \overline{x}_{jr} - \omega_{jr}(\overline{x}_{jr}; \mathbf{w}_{-r})) \right].$$

The maximum on the right hand side of the preceding expression decomposes into separate maximizations for each link $j$. We conclude that for each link $j$, we have in fact:

$$x_{jr}(\mathbf{w}) \in \arg\max_{\overline{x}_{jr} \geq 0} \left[ \alpha_{jr} \overline{x}_{jr} - \omega_{jr}(\overline{x}_{jr}; \mathbf{w}_{-r}) \right]. \tag{86}$$

Fix a link $j$. We view the users as playing a single link game at link $j$, with utility function for user $r$ given by $U_{jr}(x_{jr}) = \alpha_{jr} x_{jr}$. The preceding expression implies that (81) in Lemma 15 is satisfied, so we conclude that $\mathbf{w}$ is a Nash equilibrium for this single link game at link $j$. More precisely, we have that $(w_{j1}, \ldots, w_{jR})$ is a Nash equilibrium for the single link game at link $j$, when $R$ users with utility functions $(U_{j1}, \ldots, U_{jR})$ compete for link $j$. The maximum aggregate surplus for this link is given by:

$$\max_{\overline{x}_{j1},\ldots,\overline{x}_{jR} \geq 0} \left[ \sum_r \alpha_{jr} \overline{x}_{jr} - C_j\left(\sum_r \overline{x}_{jr}\right) \right] = \max_{\overline{f}_j \geq 0} \left[ (\max_r \alpha_{jr}) \overline{f}_j - C_j(\overline{f}_j) \right].$$

Now if $\alpha_{jr} = 0$, then from (86), the optimal choice for user $r$ is $x_{jr}(\mathbf{w}) = 0$. Thus there are two possibilities: either $\alpha_{jr} = 0$ for all $r$, in which case both the Nash equilibrium aggregate surplus and maximum aggregate surplus are zero; or $\alpha_{jr} > 0$ for at least one user $r$, in which case the maximum aggregate surplus is strictly positive, and we can apply Theorem 8 to find:

$$\sum_r \alpha_{jr} x_{jr}(\mathbf{w}) - C_j(f_j(\mathbf{w})) \geq \left(4\sqrt{2} - 5\right) \left( \max_{\overline{f}_j \geq 0} \left[ (\max_r \alpha_{jr}) \overline{f}_j - C_j(\overline{f}_j) \right] \right). \tag{87}$$

In particular, note that the preceding inequality holds for all links $j$ (since it holds trivially for those links where $\alpha_{jr} = 0$ for all $r$).

We now complete the proof of the theorem, by following the proof of Step 1 of Theorem 8. Note that we have:

$$\sum_r \boldsymbol{\alpha}_r^\top \mathbf{x}_r^S - \sum_j C_j(f_j^S) = \sum_j \left( \sum_r \alpha_{jr} x_{jr}^S - C_j(f_j^S) \right)$$
$$\leq \sum_j \max_{\overline{f}_j \geq 0} \left[ (\max_r \alpha_{jr}) \overline{f}_j - C_j(\overline{f}_j) \right]. \quad (88)$$

Expanding the definition of surplus$(\cdot, \cdot)$, we reason as follows, using (85) for the first inequality, and (88) for the second:

$$\frac{\text{surplus}(\mathbf{d}^{NE}, \mathbf{f}^{NE})}{\text{surplus}(\mathbf{d}^S, \mathbf{f}^S)} = \frac{\sum_r U_r(d_r(\mathbf{x}_r(\mathbf{w}))) - \sum_j C_j(f_j(\mathbf{w}))}{\sum_r U_r(d_r(\mathbf{x}_r^S)) - \sum_j C_j(f_j^S)}$$

$$\geq \frac{\sum_r \left( U_r(d_r(\mathbf{x}_r(\mathbf{w}))) - \boldsymbol{\alpha}_r^\top \mathbf{x}_r(\mathbf{w}) \right) + \sum_r \boldsymbol{\alpha}_r^\top \mathbf{x}_r(\mathbf{w}) - \sum_j C_j(f_j(\mathbf{w}))}{\sum_r \left( U_r(d_r(\mathbf{x}_r(\mathbf{w}))) + \boldsymbol{\alpha}_r^\top (\mathbf{x}_r^S - \mathbf{x}_r(\mathbf{w})) \right) - \sum_j C_j(f_j^S)}$$

$$= \frac{\sum_r \left( U_r(d_r(\mathbf{x}_r(\mathbf{w}))) - \boldsymbol{\alpha}_r^\top \mathbf{x}_r(\mathbf{w}) \right) + \sum_r \boldsymbol{\alpha}_r^\top \mathbf{x}_r(\mathbf{w}) - \sum_j C_j(f_j(\mathbf{w}))}{\sum_r \left( U_r(d_r(\mathbf{x}_r(\mathbf{w}))) - \boldsymbol{\alpha}_r^\top \mathbf{x}_r(\mathbf{w}) \right) + \sum_r \boldsymbol{\alpha}_r^\top \mathbf{x}_r^S - \sum_j C_j(f_j^S)}$$

$$\geq \frac{\sum_r \left( U_r(d_r(\mathbf{x}_r(\mathbf{w}))) - \boldsymbol{\alpha}_r^\top \mathbf{x}_r(\mathbf{w}) \right) + \sum_j \left( \sum_r \alpha_{jr} x_{jr}(\mathbf{w}) - C_j(f_j(\mathbf{w})) \right)}{\sum_r \left( U_r(d_r(\mathbf{x}_r(\mathbf{w}))) - \boldsymbol{\alpha}_r^\top \mathbf{x}_r(\mathbf{w}) \right) + \sum_j \max_{\overline{f}_j \geq 0} \left[ (\max_r \alpha_{jr}) \overline{f}_j - C_j(\overline{f}_j) \right]}.$$
(89)

Since $U_r(d_r(\mathbf{0})) = U_r(0) \geq 0$, by concavity of $U_r$ and the fact that $\boldsymbol{\alpha}_r \in \partial[-G_r(\mathbf{x}_r(\mathbf{w}))]$ we have:

$$U_r(d_r(\mathbf{x}_r(\mathbf{w}))) - \boldsymbol{\alpha}_r^\top \mathbf{x}_r(\mathbf{w}) \geq 0.$$

Furthermore, from (21), we have $\alpha_{jr} > p_j(f_j(\mathbf{w}))$ if $x_{jr}(\mathbf{w}) > 0$; thus $\sum_r \alpha_{jr} x_{jr}(\mathbf{w}) > f_j(\mathbf{w}) p_j(f_j(\mathbf{w})) \geq C_j(f_j(\mathbf{w}))$, where the second inequality follows by convexity (Assumption 6). This yields:

$$0 < \sum_j \left( \sum_r \alpha_{jr} x_{jr}(\mathbf{w}) - C_j(f_j(\mathbf{w})) \right) \leq \sum_j \max_{\overline{f}_j \geq 0} \left[ (\max_r \alpha_{jr}) \overline{f}_j - C_j(\overline{f}_j) \right].$$

So we conclude from relations (87) and (89) that:

$$\frac{\sum_r U_r(d_r(\mathbf{x}_r(\mathbf{w}))) - \sum_j C_j(f_j(\mathbf{w}))}{\sum_r U_r(d_r(\mathbf{x}_r^S)) - \sum_j C_j(f_j^S)} \geq \frac{\sum_j \left( \sum_r \alpha_{jr} x_{jr}(\mathbf{w}) - C_j(f_j(\mathbf{w})) \right)}{\sum_j \max_{\overline{f}_j \geq 0} \left[ (\max_r \alpha_{jr}) \overline{f}_j - C_j(\overline{f}_j) \right]} \geq 4\sqrt{2} - 5.$$

(Observe that both denominators in this chain of inequalities are nonzero.) Since $\mathbf{w}$ was assumed to be a Nash equilibrium, this completes the proof of the theorem. $\square$

The preceding theorem uses the bound on efficiency loss for a single link to establish the efficiency loss when users are price anticipating in general networks. Note that since we knew from Theorem 8 that the bound of $4\sqrt{2} - 5$ was essentially tight for single link games, and a single link is a special case of a general network, the bound $4\sqrt{2} - 5$ is also tight in this setting.

We observe that as in [12], the essential structure in the network game we consider here is that the function $U_r(d_r(\mathbf{x}_r))$ is a concave and continuous function of the vector $\mathbf{x}_r \geq 0$, and also *nondecreasing*; that is, if $x_{jr} \geq \overline{x}_{jr}$ for all $j \in J$, then $U_r(d_r(\mathbf{x}_r)) \geq U_r(d_r(\overline{\mathbf{x}}_r))$. Thus, arguing exactly as in Section 5 of [12], we can consider a more general resource allocation game where the utility to user $r$ is a concave, continuous, nondecreasing function of the vector of resources allocated, $V_r(\mathbf{x}_r)$; all the results of this section continue to hold for this more general game.

# 6 Conclusion

This paper considers a pricing mechanism where the available resources in a network are in elastic supply. For a game where users' strategies are the payments they are willing to make, we showed that the efficiency loss is no more than 34% when users are price anticipating, for the setting of a single link (Theorem 8) as well as for a network (Theorem 14).

Important questions remain regarding an extension of this work to a dynamic context. While our results suggest that manipulation of the market in a static game setting cannot lead to arbitrarily high efficiency loss, such a result does not necessarily imply users will not be able to manipulate an algorithmic implementation of this mechanism (such as those proposed in [16]). Investigation of this point is an open research topic.

Critical to any investigation of dynamics is the nature of the *information* available to the players of the pricing game. In order to compute an optimal strategic decision users need to know not only the current price level $p(f(\mathbf{w}))$, but also the total allocated rate $f(\mathbf{w})$ and the derivative of the price $p'(f(\mathbf{w}))$ (where we have assumed for simplicity that $p$ is differentiable). We postulate that the overhead of actually collecting such detailed information in a large scale communication network is quite high; in fact, in general users do not have knowledge of either the total allocated rate or the derivative of the price at the resource. This raises an important question of information availability when users respond to price signals: users may not react optimally, so what are the users' conjectures about how their strategies affect the price? Developing more detailed models for the users' response to available price information from the network is a research direction for the future.

# Acknowledgments

This work was partially supported by the National Science Foundation under a Graduate Research Fellowship and grant ECS-0312921, as well as by the Defense Advanced Research Projects Agency under the Next Generation Internet Initiative.